\documentclass[preprint]{aastex631}

\usepackage{amsmath}
\usepackage{algorithm}
\usepackage{algpseudocode}
\usepackage{CJKutf8}

\shorttitle{Detecting Faint Asteroids}
\shortauthors{Stetzler et al.}
\graphicspath{{./}{figures/}}

\newcommand{\linccfw}{LSST Interdisciplinary Network for Collaboration and Computing, Tucson, USA}
\newcommand{\diracuw}{Dept. of Astronomy \& the DiRAC Institute, University of Washington, Seattle, USA}
\newcommand{\nau}{Department of Astronomy and Planetary Science, Northern Arizona University, Flagstaff, USA}

\newcommand{\ump}{Department of Physics, University of Michigan, Ann Arbor, USA}
\newcommand{\uma}{Department of Astronomy, University of Michigan, Ann Arbor, USA}
\newcommand{\uchile}{Departamento de Astronomía, Universidad de Chile, Santiago, Chile}
\newcommand{\cfa}{Center for Astrophysics | Harvard \& Smithsonian, Cambridge, USA}
\newcommand{\lehigh}{Department of Physics, Lehigh University, Bethlehem, USA}
\newcommand{\stuttgart}{Stuttgart University of Applied Sciences, Stuttgart, Germany}
\newcommand{\byu}{Department of Physics and Astronomy, Brigham Young University, Provo, USA}
\newcommand{\APL}{Applied Physics Lab, Johns Hopkins University, Laurel, USA}
\newcommand{\ucla}{Department of Earth, Planetary and Space Sciences, University of California Los Angeles, Los Angeles, USA}
\newcommand{\carnegie}{McWilliams Center for Cosmology, Department of Physics, Carnegie Mellon University, Pittsburgh, USA}

\begin{document}

\title{An Efficient Shift-and-Stack Algorithm Applied to Detection Catalogs}

\correspondingauthor{Steven Stetzler}
\email{stevengs@uw.edu}

\author[0000-0002-7712-6678]{Steven Stetzler}
\affiliation{\diracuw}

\author[0000-0003-1996-9252]{Mario Juri\'c}
\affiliation{\diracuw}

\author[0000-0003-0743-9422]{Pedro H. Bernardinelli} 
\affiliation{\diracuw}

\author[0000-0002-1312-5529]{Dino Bektešević} 
\affiliation{\diracuw}

\author[0000-0001-7335-1715]{Colin Orion Chandler}
\affiliation{\linccfw}
\affiliation{\diracuw}
\affiliation{\nau}

\author[0000-0001-5576-8189]{Andrew J. Connolly} 
\affiliation{\linccfw}
\affiliation{\diracuw}

\author[0000-0002-8167-1767]{Fred C. Adams}
\affiliation{\ump}
\affiliation{\uma}

\author[0000-0002-5211-0020]{Cesar Fuentes}
\affiliation{\uchile}

\author[0000-0001-6942-2736]{David W. Gerdes}
\affiliation{\ump}
\affiliation{\uma}

\author[0000-0002-1139-4880]{Matthew J. Holman} 
\affiliation{\cfa}

\author[0000-0001-7737-6784]{Hsing Wen Lin (\begin{CJK*}{UTF8}{gbsn}林省文\end{CJK*})} 
\affiliation{\ump}

\author[0000-0002-2486-1118]{Larissa Markwardt}
\affiliation{\ump}

\author[0009-0005-9955-1500]{Andrew McNeill}
\affiliation{\lehigh}

\author[0000-0002-7817-3388]{Michael Mommert}
\affiliation{\stuttgart}

\author[0000-0003-4827-5049]{Kevin J. Napier}
\affiliation{\ump}

\author[0000-0001-5750-4953]{William J. Oldroyd} 
\affiliation{\nau}

\author[0000-0001-5133-6303]{Matthew J. Payne}
\affiliation{\cfa}


\author[0000-0002-9939-9976]{Andrew S. Rivkin}
\affiliation{\APL}

\author[0000-0002-6514-318X]{Luis E. Salazar-Manzano} 
\affiliation{\uma}

\author[0000-0002-0298-8089]{Hilke Schlichting}
\affiliation{\ucla}

\author[0000-0003-3145-8682]{Scott S. Sheppard}
\affiliation{\carnegie}

\author[0000-0003-4051-2003]{Dallin Spencer} 
\affiliation{\byu}

\author[0000-0001-6350-807X]{Ryder Strauss}
\affiliation{\nau}

\author[0000-0003-4580-3790]{David E. Trilling}
\affiliation{\nau}

\author[0000-0001-9859-0894]{Chadwick A. Trujillo}
\affiliation{\nau}

\begin{abstract}

The boundary of solar system object discovery lies in detecting its faintest members. However, their discovery in detection catalogs from imaging surveys is fundamentally limited by the practice of thresholding detections at signal-to-noise (SNR) $\geq 5$ to maintain catalog purity. Faint moving objects can be recovered from survey images using the shift-and-stack algorithm, which coadds pixels from multi-epoch images along a candidate trajectory. Trajectories matching real objects accumulate signal coherently, enabling high-confidence detections of very faint moving objects. Applying shift-and-stack comes with high computational cost, which scales with target object velocity, typically limiting its use to searches for slow-moving objects in the outer solar system. This work introduces a modified shift-and-stack algorithm that trades sensitivity for speedup. Our algorithm stacks low SNR detection catalogs instead of pixels, the sparsity of which enables approximations that reduce the number of stacks required. Our algorithm achieves real-world speedups of $10$--$10^3 \times$ over image-based shift-and-stack while retaining the ability to find faint objects. We validate its performance by recovering synthetic inner and outer solar system objects injected into images from the DECam Ecliptic Exploration Project (DEEP). Exploring the sensitivity--compute time trade-off of this algorithm, we find that our method achieves a speedup of $\sim30\times$ with $88\%$ of the memory usage while sacrificing $0.25$ mag in depth compared to image-based shift-and-stack. These speedups enable the broad application of shift-and-stack to large-scale imaging surveys and searches for faint inner solar system objects. We provide a reference implementation via the \texttt{find-asteroids} Python package and this URL: \url{https://github.com/stevenstetzler/find-asteroids}.

\end{abstract}


\section{Introduction} \label{sec:intro}

Discovering new solar system objects helps constrain existing and form new theories about the formation and evolution of our solar system. The boundary of discovery lies in the observation and recovery of the faintest solar system objects, probing into smaller size regimes of the different populations of solar system small bodies. Discovering and characterizing the smallest bodies in these populations helps constrain their size distribution, supporting or refuting different theories of accretion and planetary formation \citep{2005Icar..175..111B}. 

All telescope surveys are limited in their ability to find faint objects from single-epoch imaging data, due to the practice of reporting detections that appear at a signal-to-noise ratio (SNR) of $5$ or more to avoid contaminating detection catalogs with noise. Objects that may be too faint to detect confidently in single-epoch imaging can be recovered through coaddition of images across epochs \citep{Zackay_2017_II}. 

Shift-and-stack, or digital tracking, is an algorithm used to discover faint solar system objects in multi-epoch imaging of the night sky through coaddition. In the shift-and-stack algorithm, pixels from a set of multi-epoch images are first aligned along a candidate trajectory of a solar system object, typically tracking on-sky linear motion, and then coadded together. If the candidate trajectory aligns with a real object, signal from each image will accumulate coherently in coaddition, allowing a high-confidence detection of the object in a coadded ``stack'' even if the object appears at low signal-to-noise in any individual image. In stacks of 100 images, this algorithm can discover objects $10\times$ fainter in flux (2.5 magnitudes) than would otherwise be possible, making it a powerful driver for discovery of faint solar system objects. Shift-and-stack has been applied broadly to drive discovery of Trans-Neptunian objects \citep{2004AJ....128.1364B, DEEPV, DEEPVI, Fraser_2024}, main belt asteroids \citep{2015AJ....150..125H}, and near-Earth objects \citep{2014ApJ...782....1S}.

The main barrier to applying shift-and-stack is its enormous computational expense: the number of pixel stacks required for a complete search can easily exceed $10^{10}$ for small scale searches and take hours or days of compute time. Speedup can be achieved through improved parallelism, for example application of shift-and-stack on GPUs \citep{Whidden_2019, Smotherman_2021}, or through reducing the complexity of the algorithm, for example by re-using values in the stacking process \citep{Nguyen_2024}. 

Our approach to speedup is to stack low-SNR detection catalogs, which are necessarily sparser than the images from which they are derived. We utilize the work of \cite{Budavári_2017} which showed that it is possible to stack detection catalogs and recover faint objects by analyzing the relative Bayesian evidence that a set of detections are generated from an astrophysical object or a noise process. We combine this approach with and build on the work of \cite{ipol.2017.208} which outlined a method to find lines in three-dimensional point cloud data through the combination of 1) a Hough transform algorithm which roughly models the line in three dimensions and 2) linear regression which produces a refined model of the line. 

The Hough transform represents a class of template matching algorithms most often used in the field of computer vision for identifying geometric features in two-dimensional digital images. The transform was first proposed for identifying linear tracks in images of bubble chamber experiments \citep{Hough:1959qva} and later extended to identify lines and curves \citep{Hough2} as well as generalized shapes \citep{GeneralizedHough} in digital images. The central feature of these algorithms is an accumulator array whose entries correspond to the significance of a given pattern with a certain parametrization existing in an input signal. In \cite{ipol.2017.208}, the pattern is a line in three-dimensional space and the input signal are $(x, y, z)$ coordinates of point cloud data. This technique has been applied before in the context of motion tracking and remote sensing \citep{target_detection, Shuya_KASHIOKA202422.15} and exists in a broader context of signal-detection in point-cloud data \citep{Li_2023_CVPR, rs15030548}. Additionally, the work of \cite{Zhang_2024} has already shown that the method of \cite{ipol.2017.208} can be used to discover moving objects in detection catalogs. In this work, we make two modifications to the \cite{ipol.2017.208} algorithm that enable its use with low-SNR detection catalogs and connect it to the shift-and-stack algorithm.

First, we modify the Hough transform to search for two-dimensional lines in Right Ascension ($\alpha$) and Declination ($\delta$) indexed by time $t$ as an independent variable, operating on a point cloud of $(t, \alpha(t), \delta(t))$ coordinates. This alters the regression problem which the Hough transform roughly solves from total least-squares in three-dimensions to ordinary least-squares in two-dimensions, and explicitly connects our work to shift-and-stack which solves a similar regression problem \cite{Whidden_2019}. This modification behaves similarly to the Hough transform approach of Tracklet-less Heliocentric Orbit Recovery (THOR), in which clusters are identified in the residuals between the predicted and observed positions of moving objects using hypothesized orbits, roughly solving the regression problem posed by orbit fitting \citep{Moeyens_2021}. This can also be considered an extension of the moving object detection scheme based on the traditional two-dimensional Hough transform presented in \cite{DEEPIV}, which finds lines in the projected space (reduced along the temporal axis) of Right Ascension and Declination, which itself is a catalog-based complement to streak detection algorithms applied to (coadded) images \citep{2010AAS...21641511K, 2014acm..conf..570V, 2016ascl.soft05009K, VIRTANEN20161607, 10.1093/mnras/stx1565, Nir_2018}. We additionally modify the linear regression component of \cite{ipol.2017.208} to make the regression robust to outliers using the work of \cite{Rousseeuw01082004}. Incorporating robust regression makes recovery of moving objects in low-SNR catalogs feasible. It additionally makes our algorithm potentially useful in other contexts, such as moving object discovery with space-based observatories, where noise from cosmic rays can significantly contaminate the imaging and detection catalogs unless detected and masked properly \citep{Rhoads_2000, van_Dokkum_2001}. 

These modifications allow us to create a computationally efficient catalog-based shift-and-stack algorithm that trades off sensitivity to the faintest moving objects in order to attain computational speedup. This algorithm has input parameters, such as the catalog SNR, that can be tuned to achieve a desired depth or a desired speedup while controlling for the rate of false-positive detections. Through this demonstration, we show that our algorithm is an approximate shift-and-stack algorithm that bridges the gap between the first step of catalog-based moving object processing systems, making intra-night tracklets using detections from a single night of imaging \citep{kubica2007efficient, Denneau_2013, ZMODE, Holman_2018, Heinze_heliolinc}, and image-based moving object detection frameworks such as \texttt{KBMOD}, with tunable knobs that turn it from the former into the latter.

In this manuscript, we present our algorithm and validate its ability to find moving objects in both high and low-SNR detection catalogs. Section \ref{sec:algorithm} outlines our algorithm and its approach to shift-and-stack. We derive theoretical predictions of our algorithm's ability to recover moving objects in catalogs of variable SNR and derive theoretical scaling relationships of the algorithm's computational cost in terms of its input parameters. In section \ref{sec:application}, we apply the algorithm on detection catalogs derived from a Dark Energy Camera \citep[DECam;][]{DECam} imaging survey, and analyze the performance of the algorithm in terms of its ability to recover faint synthetic objects injected into the imaging dataset and the computational cost to do so. Finally, in section \ref{sec:comparison}, we make an explicit comparison between the performance of this algorithm and that of image-based shift-and-stack. 

We find that our algorithm achieves measured speedups of $10-10^3 \times$ and is able to accurately recover moving objects even in low-SNR detection catalogs. In particular, we find our algorithm achieves computational speedups of $\sim 30 \times$ with 88\% of the memory usage while sacrificing 0.25 mag in depth when compared with results from an image-based shift-and-stack search. A reference implementation of this algorithm is provided via the \texttt{find-asteroids} Python package \citep{stetzler_find_asteroids} available broadly via the Python Package Index (PyPI).\footnote{The source code for this package is available on GitHub at: \url{https://github.com/stevenstetzler/find-asteroids}}

\section{Efficient And Exhaustive Stacking of Detection Catalogs} \label{sec:algorithm}

Our algorithm applies the shift-and-stack technique to detection catalogs: lists of the observation times and on-sky positions in right ascension and declination of observed astronomical sources. The rate of motion and curvature of the on-sky position of a solar system object is dependent on the object's orbit, and typically most strongly dependent on its distance from the Earth: objects near the earth tend to move faster on the sky while those farther away move slower. Under short enough timescales, all solar system objects will appear to have linear motion on the sky. \cite{2015AJ....150..125H} notes that this time is ~1-2 hours for near earth objects, ~8 hours for main belt asteroids, and more than 24 hours for trans-Neptunian objects when observed at opposition.\footnote{These timescales are shorter at other points in a moving object's orbit where the curvature of the object's motion with respect to the Earth's motion is larger.} While shift-and-stack can be applied exhaustively to track a space of orbits, an exhaustive search over on-sky linear motion will track all orbits sufficiently over short enough timescales. This algorithm focuses on this regime: searching for linear motion in the on-sky positions of sources in the detection catalog.

In a locally-tangent plane approximation of the celestial sphere, linear on-sky motion is modeled using the equations
\begin{align}
    \alpha(t) &= (t - t_0) v_\alpha + \alpha_0 \\
    \delta(t) &= (t - t_0) v_\delta + \delta_0
\end{align}
where $v_\alpha$ is the on-sky velocity in RA, $v_\delta$ is the on-sky velocity in Dec., and $\alpha_0, \delta_0$ are the RA and Dec. at a reference epoch i.e.\ when $t = t_0$. 

If one makes a guess at the motion using a hypothesis velocity $v_\alpha^t, v_\delta^t$, the residuals of the sky position are
\begin{align}
    \alpha(t) &= (t - t_0) v_\alpha + \alpha_0 \\
    \alpha^t(t) &= (t - t_0) v_\alpha^t + \alpha_0^t \\
    \Delta \alpha = \alpha(t) - \alpha^t(t) &= (t - t_0) (v_\alpha - v_\alpha^t) + (\alpha_0 - \alpha_0^t) \\
    \delta(t) &= (t - t_0) v_\delta + \delta_0 \\
    \delta^t(t) &= (t - t_0) v_\delta^t + \delta_0^t \\
    \Delta \delta = \delta(t) - \delta^t(t) &= (t - t_0) (v_\delta - v_\delta^t) + (\delta_0 - \delta_0^t) ~.
\end{align}
From this equation, and considering just right ascension $\alpha$ (the same results will hold for $\delta$), we note that the magnitude of the residuals $\Delta \alpha$ scales linearly with the difference between the observation time $t$ and the reference epoch $t_0$ and with the difference between the trial velocity $v_\alpha^t$ and the true velocity $v_\alpha^*$. For a given moving object that has been observed multiple times and with a fixed hypothesis $v_\alpha^t$, the residuals are maximized at the time farthest from the reference epoch
\begin{align}
    \text{max}\left[ \Delta \alpha \right] &= \text{max}\left[ t - t_0\right] (v_\alpha^* - v_\alpha^t) + (\alpha_0^* - \alpha_0^t)
\end{align}
and minimized at the time closest to the reference epoch
\begin{align}
    \text{min}\left[ \Delta \alpha \right] &= \text{min}\left[ t - t_0\right] (v_\alpha^* - v_\alpha^t) + (\alpha_0^* - \alpha_0^t) ~.
\end{align}
This implies that the residuals among all of the observations of a single object are bounded in a range $\Delta x$
\begin{align}
    \Delta x &= \text{max}\left[ \Delta \alpha \right] - \text{min}\left[ \Delta \alpha \right] \\
    &= \text{max}\left[ t - t_0\right] (v_\alpha^* - v_\alpha^t) + (\alpha_0^* - \alpha_0^t) \\
    &~- \text{min}\left[ t - t_0\right] (v_\alpha^* - v_\alpha^t) - (\alpha_0^* - \alpha_0^t) \\
    &= \left(\max\left[t\right] - \min\left[t\right]\right)\left(v_\alpha^* - v_\alpha^t\right) \\
    \Delta x &= \Delta t \Delta v_\alpha
\end{align}
where $\Delta t$ is the time difference between the first and last observation of the object and $\Delta v_\alpha$ is the difference between the true and hypothesized velocity of the object. This result implies that the positional residuals for a given object are bounded within some radius $\Delta x$. $\Delta x$ will be large when the velocity hypothesis is far from the truth, say for stationary objects, and small when the velocity hypothesis is close to the truth, say for moving objects. If instead of fixing the hypothesis $v_\alpha^t$ and observing how $\Delta x$ changes we choose to fix the value of $\Delta x$ and vary the hypothesis, we find that an object's residuals will stay in a cluster of size $\Delta x$ when
\begin{align}
    \Delta v_\alpha \leq \frac{\Delta x}{\Delta t}~.
\end{align}
The same relationship is found when analyzing the residuals in declination 
\begin{align}
    \Delta v_\delta \leq \frac{\Delta x}{\Delta t}~.
\end{align}
These results show that a cluster finding algorithm can be applied to the space of projected positions using the clustering radius $\Delta x$ to recover moving objects with true velocities within $\Delta v_\alpha, \Delta v_\delta$ of a hypothesis $v_\alpha^t, v_\delta^t$.\footnote{Any cluster finding algorithm which utilizes a tunable clustering radius parameter can be used here, including the popular \texttt{DBSCAN} algorithm \citep{DBSCAN}.} These results also provide a method to construct the set of velocities that must be hypothesized to perform a complete blind search for a target population of moving objects: simply test all on-sky velocities between $v_{\textrm{min}}$ and $v_{\textrm{max}}$ which differ in their $\alpha$ and $\delta$ components by no more than $\Delta v_{\alpha}$ and $\Delta v_{\delta}$. For a given value of $\Delta x$, this results in a set of velocities $\mathcal{V}$ of size
\begin{align} \label{eq:v_size}
    \begin{split}
        \vert \mathcal{V} \vert &= \frac{v_{\mathrm{max}, \alpha} - v_{\mathrm{min},\alpha}}{\Delta v_{\alpha}}  \frac{v_{\mathrm{max}, \delta} - v_{\mathrm{min},\delta}}{\Delta v_{\delta}} \\
        &= \left(\frac{\Delta t \Delta v}{\Delta x} \right)^2
    \end{split}
\end{align}
where $\Delta v = v_{\textrm{max}} - v_{\textrm{min}}$ is assumed to be the same for $\alpha$ and $\delta$. The complete blind search can be thought of as exploring the set of velocity hypotheses that bound the residuals of objects from the target population within a threshold $\Delta x$. 

Once a cluster is found in the projected space, its exact velocity is not known except that it is plausibly within $\Delta x / \Delta t$ of the initial hypothesis velocity. The true velocity of the object can be recovered by repeating the blind search procedure on the clustered data alone and with $\Delta x$ set to a smaller value, perhaps a value that would be reflective of the expected positional residuals given the object's expected brightness. However, if the cluster is relatively pure, meaning it mostly contains detections from a single moving object, a data-driven approach will suffice in which case the true velocity of the object can be recovered via multivariate regression. The multivariate regression problem with $N$ observations is cast as
\begin{align}
    Y &= B^T X + c + \epsilon
\end{align}
where $Y \in \mathbb{R}^{N \times q}$ are the response variables, $X \in \mathbb{R}^{N \times p}$ are the independent variables, $B \in \mathbb{R}^{p \times q}$ is the slope matrix, $c \in \mathbb{R}^{q}$ is the intercept vector and $\epsilon \sim \mathcal{N}(0, \Sigma_\epsilon)$ is a random variable representing errors in observations and is modeled as a q-dimensional normal distribution with zero-mean. 

The slope, intercept, and covariance of the errors are estimated from the mean and covariance matrices of the response and independent variables. If we partition the mean $\mu$ and covariance $\Sigma$ according to $X$ and $Y$
\begin{align}
    \mu &= \begin{bmatrix}
        \mu_X \\
        \mu_Y
    \end{bmatrix} \Sigma = \begin{bmatrix}
        \Sigma_{XX} & \Sigma_{XY} \\
        \Sigma_{YX} & \Sigma_{YY}
    \end{bmatrix}
\end{align}
then the slope matrix is estimated from the sample mean $\hat{\mu}$ and sample covariance matrix $\hat{\Sigma}$ as
\begin{align}
    B &= \hat{\Sigma}_{XX}^{-1} \hat{\Sigma}_{XY}
\end{align}
while the intercept vector is
\begin{align}
    c &= \hat{\mu}_Y - B^T\hat{\mu}_{X}
\end{align}
and the covariance of the errors is
\begin{align}
    \hat{\Sigma}_\epsilon &= \hat{\Sigma}_{YY} - B^T \hat{\Sigma}_{XX} B ~.
\end{align}
Once the velocity of the moving object is found via regression, the object can be considered recovered. 

The sample mean and covariance matrix are sensitive to the inclusion of outliers. If the cluster contains noise detections or contaminating detections from artifacts, then the recovered trajectory can be biased away from the truth. Multivariate regression can still be used if made robust to the inclusion of noise. In the prior formulation, a robust estimator of the mean and covariance of the data will provide a robust estimate of the regression parameters. The Minimum Covariance Determinant (MCD) provides such a method for robust determination of the mean and covariance of a dataset \citep{MCD, Rousseeuw01082004}. This method enumerates over subsets of the dataset of size $ n/2 \leq h < n$ to find a subset whose covariance determinant is minimized. The mean and covariance of this subset are reported as the robust estimate of the whole sample mean and sample covariance. The subset with the minimum covariance determinant achieves the minimal scatter about a central point i.e. this subset excludes outliers that may be present in the full dataset. When setting $h = n/2$, this method will be robust up to 50\% noise contamination in the dataset, achieving the best possible break down point of any method. 

\begin{figure}[h]
    \centering
    \includegraphics[width=\linewidth]{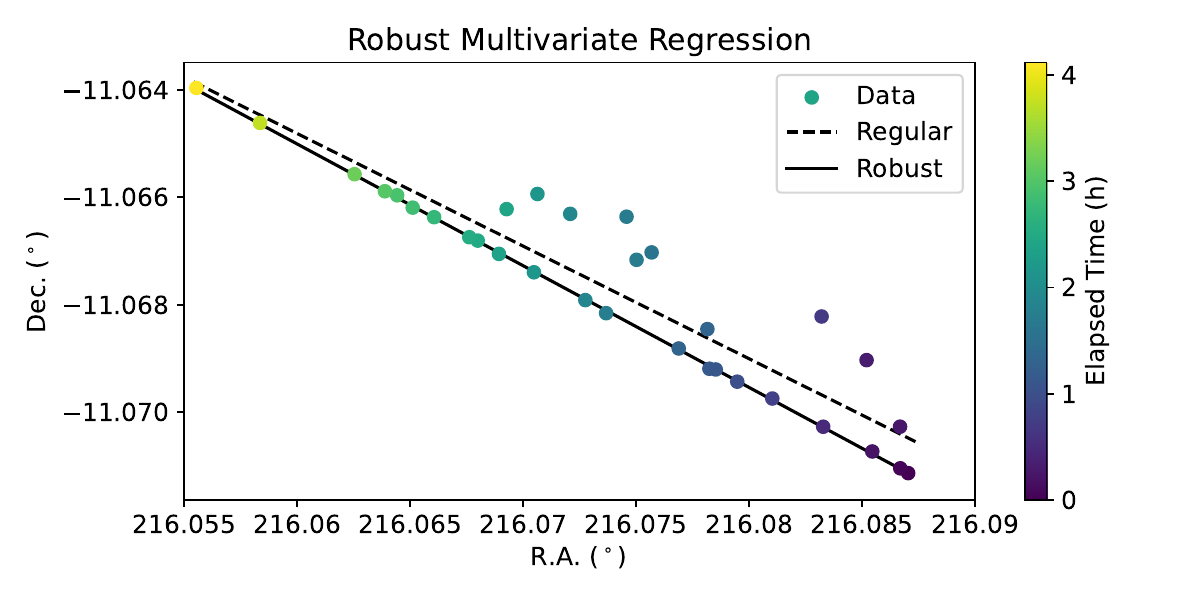}
    \caption{The on-sky coordinates of a moving object with respect to time. The dashed black line represents a trajectory estimated using least squares multivariate regression, which does not predict the trajectory accurately due to the presence of outliers in the dataset. The solid black line represents a trajectory estimated using robust multivariate regression utilizing the Minimum Covariance Determinant (MCD), which accurately predicts the moving object trajectory in the presence of outliers.}
    \label{fig:regression}
\end{figure}

\cite{Rousseeuw01081999} introduces the FAST-MCD, which enables fast iteration over subsets of large datasets to find either the exact MCD or an approximation of it. An implementation of FAST-MCD is provided in \texttt{scikit-learn}, enabling its use in Python. A Python implementation of the FAST-MCD for robust multivariate regression is provided by us via the \texttt{mcd-regression} Python package \citep{stetzler_mcd_regression}.\footnote{Additionally available via GitHub at \url{https://github.com/stevenstetzler/mcd-regression}} Figure ~\ref{fig:regression} visualize how the technique of robust multivariate regression can be used to recover a moving object's trajectory in the presence of noise.

\subsection{The Algorithm}

Based on these observations, our algorithm follows, which is in essence the iterative Hough transform algorithm presented by \cite{ipol.2017.208}. Take as input a detection catalog
\begin{align}
    \mathcal{C} &= \begin{bmatrix}
        \alpha_i & \delta_i & t_i
    \end{bmatrix}_{i=1}^{n}
\end{align}
a set of hypothesis velocities 
\begin{align}
    \mathcal{V} &= \begin{bmatrix}
        v_\alpha^j & v_\delta^j
    \end{bmatrix}_{j=1}^{m}
\end{align}
a clustering radius $\Delta x$ and a minimum value for the number of detections we expect from each moving object $N_{\mathrm{min}}$. 

\begin{enumerate}
    \item Project: First, project the detection catalog to the reference epoch, which we choose to be the minimum of all of the observation times contained in the detection catalog. 
    \item Cluster: Next, for each hypothesis velocity, construct a two dimensional histogram of the projected positions using a resolution of $\Delta x$. Consider the set of histograms jointly as a three dimensional histogram. 
    \item Find clusters: In a while loop, identify the location of the maximum of the histogram. If the maximum is $< N_{\mathrm{min}}$, break to the next stage. Otherwise, identify the points in the detection catalog that contributed to that maximum. These points form a cluster. Remove those points from the histogram. Continue until the loop is broken.
    \item Refine: For each cluster produced, perform multivariate regression with the MCD to fit a linear trajectory to the points recovered.
    \item Gather: Gather points from the original detection catalog within $\Delta x$ of the refined line. 
    \item Optional: Repeat steps (4) and (5) until convergence i.e.~the set of gathered points does not change in each iteration. 
\end{enumerate}

\subsection{Setting the algorithm parameters}

It remains to be addressed how one should set the parameters of the algorithm: $\Delta x$ and $N_{\mathrm{min}}$. Both are related to the expected purity of the clusters found by the algorithm. If the input detection catalog is already pure with no false positive detections, the value of $\Delta x$ should be set based on the sky density of astronomical sources in the catalog. If $\Delta x$ is set too large, then one will expect significant contamination in the space of projected positions from stationary sources as well as other moving objects. If the input catalog is constructed at low-SNR, meaning some (or even most) of the detections in the catalog are spurious, then the value of $\Delta x$ should be set to control the purity of the resulting cluster in the algorithm. A larger $\Delta x$ will accumulate more noise detections while a smaller $\Delta x$ will accumulate fewer noise detections in a single cluster. 

The value of $\Delta x$ can be determined by reasoning about the number of noise peaks $N_{\mathrm{noise}}$ and real object peaks $N_{\mathrm{signal}}$ expected to land in a bin of size $\Delta x^2$. The best value to choose for $\Delta x$ maximizes the ratio $\mathbb{E}[N_{\mathrm{signal}} / N_{\mathrm{noise}}]$, which is akin to a likelihood ratio. If $N_{\mathrm{signal}} / N_{\mathrm{noise}} > 1$, then the true moving object can be distinguished from the noise peaks included. If $N_{\mathrm{signal}} / N_{\mathrm{noise}} > 1$, then the contamination rate $N_{\mathrm{noise}} / (N_{\mathrm{signal}} + N_{\mathrm{noise}}) < 0.5$ and we expect the MCD multivariate regression technique to recover the object. 

Assuming the value of $\Delta x$ is chosen to maximize signal relative to noise, then the value of $N_{\mathrm{min}}$ determines the number of false positives produced by the algorithm. It can be chosen so that 
\begin{align}
    P(N_{\mathrm{noise}} \geq N_{\mathrm{min}} ) \leq \epsilon
\end{align}
where $\epsilon$ represents an admissible false positive rate of detected objects. Under the algorithm considered, a reasonable value of $\epsilon$ is 
\begin{align}
    \epsilon &= \frac{N_{\mathrm{false}}}{N_{\mathrm{stacks}}}
\end{align}
where $N_{\mathrm{false}}$ is the maximum number of false results produced by the algorithm and $N_{\mathrm{stacks}}$ is the number of independent stacks performed. If we assume the trial velocities and stack bins are independent, then the number of stacks is
\begin{align} \label{eq:N_stacks}
    \begin{split}
        N_{\mathrm{stacks}} &= |\mathcal{V}| \frac{\Omega}  {\Delta x^2} \\
        &= \frac{\Delta t^2 \Delta v^2 \Omega}{\Delta x^4}    
    \end{split}
\end{align}
using Eq.~\ref{eq:v_size} to substitute in $|\mathcal{V}|$ and where $\Omega$ is the solid angle of sky searched. What remains is to specify the distributions of $N_{\mathrm{noise}}$ and $N_{\mathrm{signal}}$.

The distribution of noise peaks has been considered in detail by \cite{Budavári_2017} which specifies the spatial distribution of noise peaks as a function of the detection statistic. Assuming the detection process uses a Gaussian window function (i.e. PSF detection) and the sky noise is white, the surface density of noise peaks produced by the detection process is (using \cite{Budavári_2017} Eq.~40 and 46)
\begin{align}
    n_{\text{pk}} &= \frac{\exp(-z^2/2)}{2 \pi^2 a^2} \left(\frac{\sqrt{2}}{4} z + B\left(\frac{\sqrt{2}}{2} z, 1\right)\left[\frac{1}{2} + \frac{3}{4} z\right] + B\left(\frac{\sqrt{2}}{2} z , 2\right)\right)
\end{align}
where $z$ is the value of the detection statistic, $a$ is the PSF size, and
\begin{align}    
    B(s, b) &\equiv \frac{\pi}{b} \exp{\left(\frac{s^2}{2b}\right)}\left[ 1 + \text{erf}\left( \frac{s}{\sqrt{2b}}\right)\right]~.
\end{align}
The cumulative number of peaks above threshold $\nu_{\mathrm{th}}$ is then
\begin{align} \label{eqn:n_pk_above}
    \lambda(\nu_{\mathrm{th}}) &= \int_{\nu_{\mathrm{th}}}^{\infty} n_{\text{pk}}(z) dz~.
\end{align}
The spatial distribution of peaks is a Poisson point process, meaning the number of noise detections $N_{\mathrm{noise}}$ is Poisson distributed
\begin{align}
    N_{\mathrm{noise}} \sim \text{Pois}\left( \Lambda \right)
\end{align}
with a rate $\Lambda$ equal to
\begin{align}
    \Lambda &= \int_A \lambda(\nu_{\mathrm{th}}) dA
\end{align}
where $A$ represents a spatial region of interest. The probability of having $N_{\mathrm{noise}} = k$ noise peaks in a square region with area $\Delta x^2$ (in units of the PSF size $a$) is
\begin{align}
    P(N_{\mathrm{noise}} = k) &= \frac{\Lambda^k e^{-\Lambda}}{k!} \\ &=\frac{(\lambda(\nu_{\mathrm{th}})\Delta x^2)^k e^{-\lambda(\nu_{\mathrm{th}})\Delta x^2}}{k!} ~.
\end{align}
If we consider multiple epochs $i$ with different values of the PSF size $a_i$, the number of noise peaks above a threshold $\nu_{\mathrm{th}}$ within a stacked bin of size $(\Delta x / a_i)^2$ is
\begin{align}
    N_{\mathrm{noise}} &\sim \sum_{i} \text{Pois}(\Lambda_i) \\
    &\sim \text{Pois}\left(\sum_{i} \Lambda_i\right) \\
    &\sim \text{Pois}\left(\sum_{i} \lambda(\nu_{\mathrm{th}})(\Delta x / a_i)^2\right) ~.
\end{align}

The number of real object detections $N_{\mathrm{signal}}$ depends on the probability that an object is observed with an above-threshold flux and the probability that its observed position is within the spatial region of interest. The observed flux of a source is normally distributed
\begin{align}
    \hat{f} \sim \mathcal{N}\left(f, \sigma_f^2 \right)
\end{align}
where $\hat{f}$ and $f$ are the observed and true fluxes of an object, and $\sigma_f$ is the associated uncertainty with the flux measurement. The object is detected above a threshold $f_{\mathrm{th}}$ with probability
\begin{align}
    P(\hat{f} \geq f_{\mathrm{th}} | f) &= \frac{1}{2} \text{erfc}\left( \frac{f_{\mathrm{th}} - f}{\sigma\sqrt{2}}\right) ~.
\end{align}
Expressed in term of the signal-to-noise $\nu = f / \sigma_f$ we find
\begin{align}
    P(\hat{\nu} \geq \nu_{\mathrm{th}} | \nu) &= \frac{1}{2} \text{erfc}\left( \frac{\nu_{\mathrm{th}} - \nu}{\sqrt{2}}\right) ~.
\end{align}
The observed position $x$ of the object at a single epoch is also normally distributed
\begin{align}
    \hat{\alpha} &\sim \mathcal{N}\left(\alpha, \sigma_\alpha^2 \right)  \\
    \hat{\delta} &\sim \mathcal{N}\left(\delta, \sigma_\delta^2 \right)
\end{align}
where $\sigma_\alpha$ and $\sigma_\delta$ are the positional uncertainties. The positional uncertainties depend on the PSF size $a$ and the signal-to-noise of the object $\nu$. \cite{Portillo_2020} provides a derivation of the positional uncertainties (Eq.~14):
\begin{align}\label{eqn:pos_uncertainty}
    \sigma_\alpha^2 = \sigma_\delta^2 &= 2 a^2 \left(\frac{f^2}{\sigma_f^2}\right)^{-1} \\
    &= 2 \frac{a^2}{\nu^2}
\end{align}
where $\nu$ is the object signal-to-noise. The probability that an object is observed within $\Delta x$ of the true position $\alpha, \delta$ is then
\begin{align}
    P(|\hat{\alpha} - \alpha| \leq \Delta x/2) &= 1 - 2 P( \hat{\alpha} - \alpha  \leq -\Delta x/ 2) \\ 
    &= \text{erf}\left(\frac{\Delta x}{\sigma_\alpha2 \sqrt{2}}\right).
\end{align}
Substituting the relationship between $\sigma_\alpha$ and signal-to-noise $\nu$ we find
\begin{align}
     P(|\hat{\alpha} - \alpha| \leq \Delta x/2) &= \text{erf}\left(\frac{\Delta x \nu}{4 a}\right) ~.
\end{align}
The probability that both $\alpha$ and $\delta$ are within $\Delta x$ is then
\begin{align}
    P(|\hat{\alpha} - \alpha| \leq \Delta x/2, |\hat{\delta} - \delta| \leq \Delta x/2) &= \text{erf}\left(\frac{\Delta x \nu}{4 a}\right)^2 ~.
\end{align}

If we assume the trial velocity $v_\alpha^t, v_\delta^t$ exactly matches the true moving object trajectory, such that $\alpha(t) = \alpha^t(t)$ and $\delta(t) = \delta^t(t)$, then the probability of the detection from a single epoch landing in a bin of size $\Delta x^2$ is
\begin{align}
    &P(\hat{\nu}_i \geq \nu_{\mathrm{th}}, |\alpha^t(t) - \hat{\alpha}_i | \leq \Delta x, |\delta^t(t) - \hat{\delta}_i | \leq \Delta x) \\
    &= \frac{1}{2} \text{erfc}\left(\frac{\nu_{\mathrm{th}} - \nu}{\sqrt{2}}\right)\text{erf}\left(\frac{\Delta x \nu}{4a}\right)^2 ~.
\end{align}
Across many epochs, this probability is repeated with a per-epoch PSF $a_i$. $N_{\mathrm{signal}}$ is the number of times that this outcome occurs and is the result of several Bernoulli trials with per-epoch rates. $N_{\mathrm{signal}}$ is then Poisson Binomial distributed. Making the simplifying assumption that the per-epoch PSF is constant and equal to $a$, then $N_{\mathrm{signal}}$ follows a Binomial distribution:
\begin{align}
    P(N_{\mathrm{signal}} = \vert \mathcal{D} \vert) &= \prod_{i \in \mathcal{D}} p_i \prod_{i \in \mathcal{D}^c} (1 - p_i)
\end{align}
where $p_i$ is the probability of detection in a single epoch and $\mathcal{D}$ ($\mathcal{D}^c$) is the set of epochs the object is (not) detected in. 

\begin{figure*}[h]
    \centering
    \includegraphics[width=\linewidth]{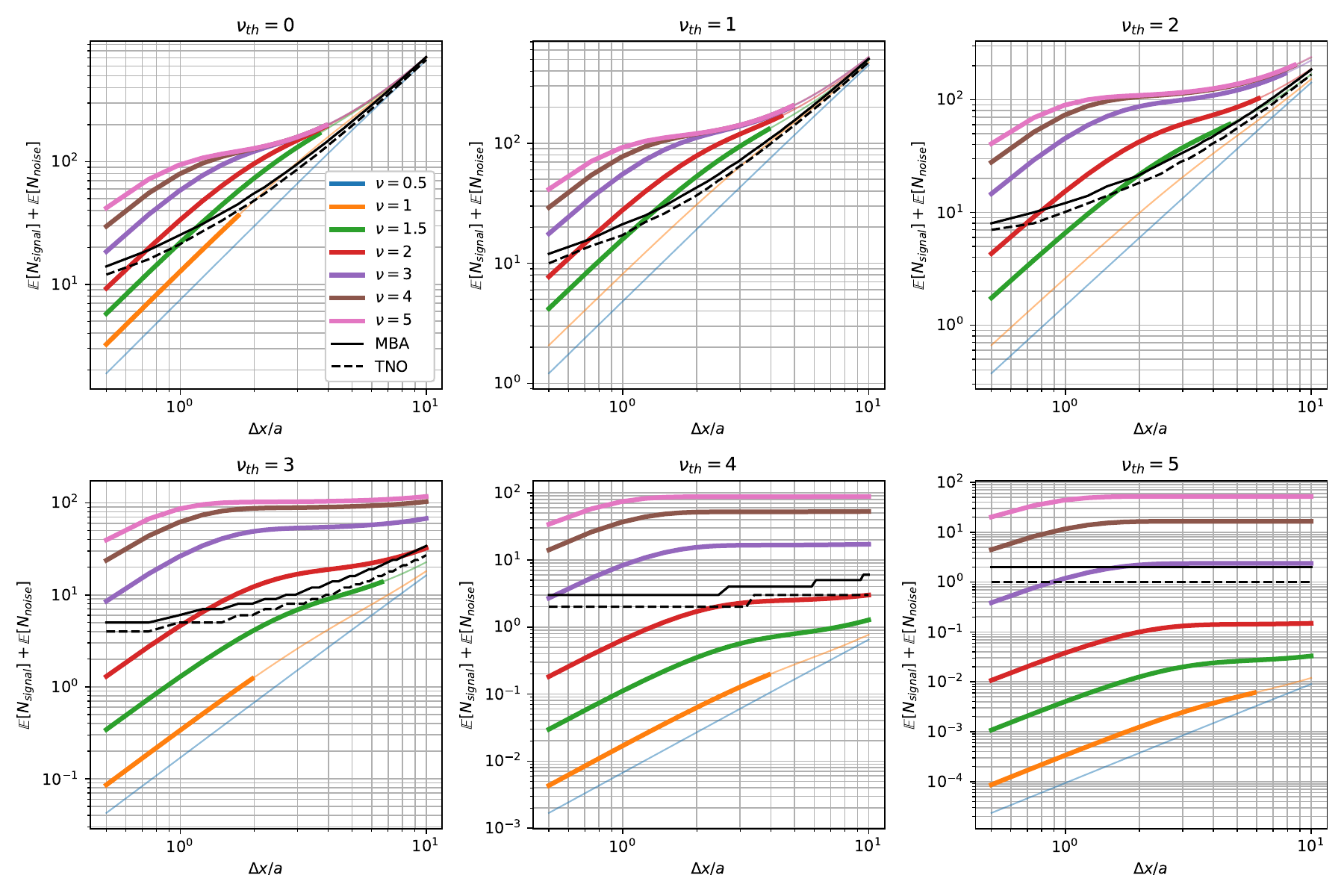}
    \caption{The expected number of detection counts in a bin of size $\Delta x$ for objects of different SNR $\nu$ found in catalogs of different SNR thresholds $\nu_{\mathrm{th}}$. Thick (thin) colored lines approximately indicate when the expected number of signal detections will be larger (less) than the expected number of noise detections i.e. when $\mathbb{E}[N_{\mathrm{signal}}] / \mathbb{E}[N_{\mathrm{noise}}] > 1$ (or $< 1$), indicating when a single object of the given SNR $\nu$ is distinguishable from noise in a stack of the detection catalog that corresponds to the given object's trajectory. Black solid (dashed) lines indicate values of $N_{\mathrm{min}}$ that produce a false positive rate of $\epsilon = 100/N_{\mathrm{stacks}}$ in an example MBA (TNO) search covering $\Omega = 162 \text{ arcmin}^2$ with $a = 1 \text{ arcsec}$ seeing and a $\Delta t = 4$ hour time baseline. Colored lines lying in regions of the parameter space $N_{min} = \mathbb{E}[N_{\mathrm{signal}}] + \mathbb{E}[N_{\mathrm{noise}}]$ and $\Delta x$ above (below) the black lines represent when objects of the given SNR $\nu$ are (un)likely to appear in the results of a search before producing 100 results corresponding to stacks of noise detections.}
    \label{fig:choose_dx_Nmin}
\end{figure*}

The distributions of $N_{\mathrm{signal}}$ and $N_{\mathrm{noise}}$ depend on the SNR of the object, the SNR threshold used to derive the catalog, and the PSF width at each epoch. Figure \ref{fig:choose_dx_Nmin} provides a visual guide for choosing reasonable values of $\Delta x$ and $N_{\mathrm{min}}$ based on the expected values of $N_{\mathrm{signal}}$ and $N_{\mathrm{noise}}$. Each panel in the figure plots the expected number of counts $\mathbb{E}[N_{\mathrm{signal}}] + \mathbb{E}[N_{\mathrm{noise}}]$ as a function of the aperture size $\Delta x$ for objects of varying SNR $\nu$ found in a detection catalog of varying SNR thresholds $\nu_{\mathrm{th}}$. For each value of $\nu$, the plot indicates when $\mathbb{E}[N_{\mathrm{signal}}] / \mathbb{E}[N_{\mathrm{noise}}]$ (used as an approximation of the harder to compute $\mathbb{E}[N_{\mathrm{signal}}/N_{\mathrm{noise}}]$) is greater than or less than $1$, indicating that an object is recoverable or not using the MCD regression technique. Additionally, a predicted value for $N_{\mathrm{min}}$ that is expected to produce $100/N_{\mathrm{stacks}}$ false detection candidates for an MBA and TNO search is overplotted, representing the ``noise floor'' for a given search; setting $N_{\mathrm{min}}$ lower than these values can recover fainter objects at the expense of increasing numbers of false positives. This figure indicates which objects are expected to be detectable above the noise floor Notably, this figure illustrates that not all objects of a given SNR $\nu$ are recoverable over all aperture sizes (or any in the case of $\nu = 0.5$). This is due to the scaling of the positional uncertainties with SNR, which is expected to diverge as $\nu \rightarrow0$; at some or all aperture sizes, the rate with which noise detections are accumulated surpasses the real object detection rate. 

\subsection{Algorithmic Scaling}

An image-based shift-and-stack algorithm has algorithmic complexity in number of operations that scales as 
\begin{align}
    \mathcal{O}(N_{\mathrm{t}}  N_{\mathrm{pixels}}  \vert \mathcal{V} \vert)
\end{align}
or 
\begin{align}\label{eqn:sns_ops}
    \mathcal{O}\left(N_{\mathrm{t}}  \frac{\Omega}{p_s^2}  \vert \mathcal{V} \vert\right)
\end{align}
where $N_{\mathrm{t}}$ is the number of epochs/number of images stacked, $N_{\mathrm{pixels}}$ is the number of pixels per image, $\vert \mathcal{V} \vert$ is the size of the velocity set searched, $\Omega$ is the solid angle subtended by a single image, and $p_s$ is the image pixel scale. This is due to the fact that two operations (shift + add) are performed on every pixel in the set of provided input images for each direction searched. The memory usage of this algorithm scales as 
\begin{align}\label{eqn:sns_mem}
    \mathcal{O}(N_{\mathrm{t}}  N_{\mathrm{pixels}})
\end{align}
since all images need to be stored in memory during the procedure. 

In our algorithm, $N_{\mathrm{pixels}}$ is replaced with $N_{\mathrm{peaks}}$ where $N_{\mathrm{peaks}}$ is the number of detection peaks per image. Assuming the detections are dominated by noise, then $N_{\mathrm{peaks}}$ is computed relative to $N_{\mathrm{pixels}}$ using Eq.~\ref{eqn:n_pk_above} as
\begin{align}
    N_{\mathrm{peaks}} &= \lambda(\nu_{\mathrm{th}})  \frac{N_{\mathrm{pixels}}}{a_p^2} \\
    &= \lambda(\nu_{\mathrm{th}})  \frac{N_{\mathrm{pixels}} p_s^2}{a^2} \\
    &= \lambda(\nu_{\mathrm{th}}) \frac{\Omega}{a^2}
\end{align}
where $\Omega$ is the solid angle subtended by a single image, $p_{s}$ is the pixel scale, $a_{p}$ is the PSF-width in pixels and $a$ is the PSF-width in angular units. The process of source detection scales as
\begin{align}
    \mathcal{O}\left( N_{\mathrm{t}}  N_{\mathrm{pixels}} \right)
\end{align}
in operations and
\begin{align}
    \mathcal{O}\left(N_{\mathrm{pixels}} \right)
\end{align}
in memory assuming the detection process is applied sequentially to the input images. The complexity, accounting for steps 1 (Project) and 2 (Cluster) of our algorithm, is then
\begin{align}
    \mathcal{O}\left( N_{\mathrm{t}}  \lambda(\nu_{\mathrm{th}}) \frac{\Omega}{a^2}  \vert \mathcal{V} \vert \right)
\end{align}
in operations, while the memory scaling is
\begin{align}
    \mathcal{O}\left( N_{\mathrm{t}}  \lambda(\nu_{\mathrm{th}}) \frac{\Omega}{a^2}  \vert \mathcal{V} \vert + \frac{\Omega}{\Delta x^2}  \vert \mathcal{V} \vert \right)
\end{align}
since the projected positions of the detection catalog are stored in memory in addition to the Hough space. Step 3 (Find Clusters) scales in operations as
\begin{align}
    \mathcal{O} \left( N_{\mathrm{results}}\frac{\Omega}{\Delta x^2}  \vert \mathcal{V} \vert \right)
\end{align} 
since it involves finding a maximum entry in the Hough space of this same size. The total complexity of the catalog-based shift-and-stack approach is then
\begin{align}\label{eqn:hough_ops}
    &\mathcal{O}\left(N_{\mathrm{t}}  N_{\mathrm{pixels}} + N_{\mathrm{t}}  \lambda(\nu_{\mathrm{th}}) \frac{\Omega}{a^2}  \vert \mathcal{V} \vert + N_{\mathrm{results}}\frac{\Omega}{\Delta x^2}  \vert \mathcal{V} \vert \right)
\end{align}
in operations and
\begin{align}\label{eqn:hough_mem}
    \mathcal{O}\left( N_{\mathrm{pixels}} + N_{\mathrm{t}}  \lambda(\nu_{\mathrm{th}}) \frac{\Omega}{a^2}  \vert \mathcal{V} \vert + \frac{\Omega}{\Delta x^2}  \vert \mathcal{V} \vert \right)
    \end{align}
in memory accounting for all steps.

Taking only the terms in Eqns.~\ref{eqn:sns_ops}, \ref{eqn:sns_mem}, \ref{eqn:hough_ops}, and \ref{eqn:hough_mem} which scale with the velocity set $\vert \mathcal{V} \vert$ and excluding result finding (i.e. $N_{\mathrm{results}} = 0$), we can calculate the speedup that our algorithm (2) attains relative to an image-based approach (1) in the shift-and-stack procedure itself:
\begin{align}
    s &= \frac{N_{\mathrm{t}}  \frac{\Omega}{p_s^2}  \vert \mathcal{V}_1 \vert}{N_{\mathrm{t}}  \lambda(\nu_{\mathrm{th}}) \frac{\Omega}{a^2}  \vert \mathcal{V}_2 \vert} \\
    &= \frac{a^2  \vert \mathcal{V}_1 \vert}{\lambda(\nu_{\mathrm{th}}) p_s^2  \vert \mathcal{V}_2 \vert} \\
    &= \frac{a_p^2 \vert \mathcal{V}_1 \vert}{\lambda(\nu_{\mathrm{th}})  \vert \mathcal{V}_2 \vert}
\end{align}
while the relative memory usage is
\begin{align}
    r &= \frac{N_{\mathrm{t}}  \lambda(\nu_{\mathrm{th}}) \frac{\Omega}{a^2}  \vert \mathcal{V}_2 \vert +  \frac{\Omega}{\Delta x_2^2}  \vert \mathcal{V}_2 \vert}{N_{\mathrm{t}}  \frac{\Omega}{p_s^2}} \\
    &= \vert \mathcal{V}_2 \vert \left( \lambda(\nu_{\mathrm{th}}) \frac{p_s^2}{a^2} + \frac{p_s^2}{N_{\mathrm{t}} \Delta x_2^2}\right) ~.
\end{align}
Assuming the same velocity range $\Delta v$ and time span $\Delta t$ is searched, and the two searches differ by $\Delta x$, then substituting values of $\left\vert \mathcal{V} \right\vert$ using Eq.~\ref{eq:v_size} we find the speedup and relative memory usage are
\begin{align}\label{eqn:scaling}
    s &= \frac{a_p^2}{\lambda(\nu_{\mathrm{th}})} \left( \frac{\Delta x_2}{\Delta x_1}\right)^2 \\
    r &= \left(\frac{\Delta v \Delta t}{\Delta x_2} \right)^2 p_s^2 \left(  \frac{\lambda(\nu_{\mathrm{th}})}{a^2} + \frac{1}{N_{\mathrm{t}} \Delta x_2^2}\right)~.
\end{align}

Our algorithm is efficient in the sense that it performs fewer effective stacks relative to an image-based search, due to the scaling of speedup with the value of $\Delta x^2$. Assuming the PSF width is $a_p = 5$ pixels ($1$ arcsec seeing with a 0.2 arcsec pixel scale), $\Delta x_1 = 1 a$ (1 PSF width) for an image-based search and $\Delta x_2 = 5 a$ (5 PSF widths) for the catalog-based search, then at $\nu_{\mathrm{th}} = 2$, the predicted speedup is $\approx 5 \times 10^4$. As implemented, the greatest weakness of our algorithm is in its memory usage. Since all search directions are considered jointly and the projected positions of all detections are stored at once, the memory footprint can quickly outgrow the capacity of most machines at low-SNR and low values of $\Delta x$. 

\section{Application}\label{sec:application}

In this section, we describe searches for faint moving objects in images from the Dark Energy Camera (DECam). We perform searches for both slow-moving Trans Neptunian Objects (TNOs) and faster-moving Main Belt Asteroids (MBAs). We analyze the performance of the algorithm as a function of the SNR of the catalog used in terms of its ability to recover implanted fake objects and the computation time and memory needed to perform the search. 

\subsection{Description of Data} \label{sec:data}

The data used are a single night of data from the DECam Ecliptic Exploration Project \citep[DEEP;][]{Trilling_2024, DEEPII}. The DEEP observing strategy involves tiling the sky around the invariable plane of the solar system using a long-stare observing strategy across multiple years (2019-2022). Each night of the survey comprises of one or more long-stare pointing---sequences of 120 second VR-band exposures taken with DECam over a period of 2-4 hours. The data we test our algorithm with are selected from the third night of the survey---April 3, 2019--and the long-stare observation of one field of the survey---A0c---which includes 104 120-second VR-band exposures.\footnote{\cite{Trilling_2024} outlines the nomenclature used for field naming: A refers to the A semester, A0 to a patch which tracks the year-to-year motion of TNOs, and A0c to a single field observed in that patch.} 

Data are processed using the LSST Science Pipelines \citep[LSP;][]{Bosch_2018, Bosch_2019} version \texttt{w\_2024\_09}.\footnote{The LSST Science Pipelines are available freely at \url{pipelines.lsst.io}.} Raw science images, VR-band dome flat field images, and bias images are downloaded from the NOIRLab archive.\footnote{\url{https://astroarchive.noirlab.edu/}} Bias and flat field images are processed through the LSP calibration product pipeline (\texttt{cp\_pipe}) for DECam. The output of this pipeline are per-detector nightly stacked flat and bias master calibration files which are applied to the science images. 

The science images are processed using the LSP data release product pipeline (\texttt{drp\_pipe}). Each detector is processed individually including defect masking, application of the master calibrations, cosmic ray detection and repair, PSF model fitting to bright stars, background subtraction, source detection, fitting of a world coordinate system (WCS) by crossmatching detected sources to the Gaia DR3 \citep{Gaia_2016, Gaia_2023} catalog, and photometric zero point fitting by crossmatching detected sources to the Pan-STARRS1 catalog \citep{ps1}. The photometric zero point is fit using r-band measurements from Pan-STARRS1, neglecting a color term correction for the VR-band images. The VR-band overlaps well with the r-band, meaning the resulting photometry will be approximately correct. 

\begin{figure}[h]
    \centering
    \includegraphics[width=\linewidth]{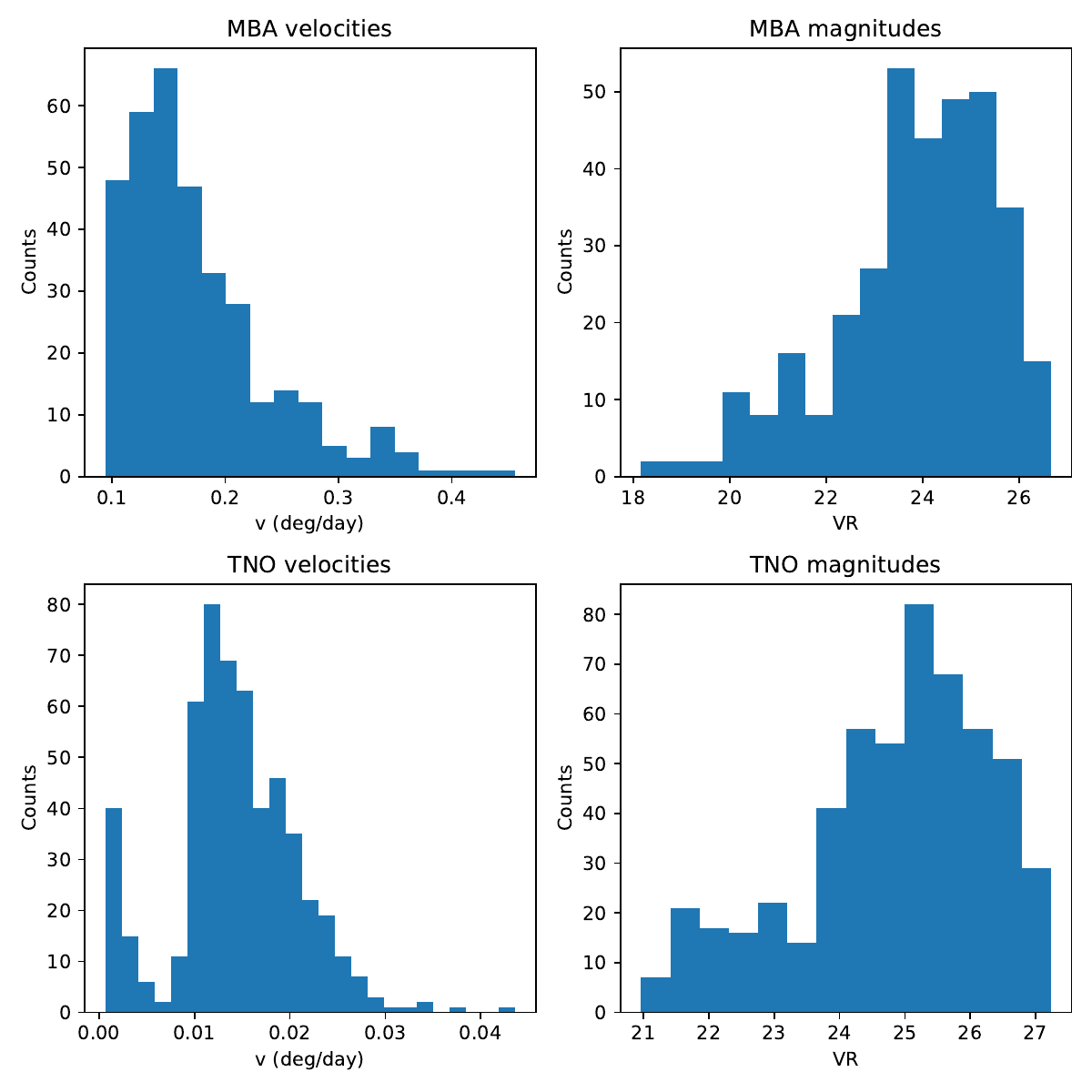}
    \caption{The distribution of velocities and magnitudes of the injected synthetic MBAs and TNOs.}
    \label{fig:fakes}
\end{figure}

We inject a population of synthetic trans-Neptunian objects as well as closer objects, representing both the main belt asteroids and the Jupiter Trojans, in the data. The TNO population is identical to that of \cite{DEEPIII}, with three subpopulations that span the full range of orbits expected in the outer Solar System. The asteroid population has their semi-major axes uniformly distributed between 2 and 6 au, their eccentricities between 0 and 0.4, and their inclinations sinusoidally distributed between 0 and 50 deg (as done for a subset of the TNO population in \cite{DEEPIII}). The final three parameters, longitude of ascending node, argument of perihelion, and mean anomaly, are uniform between $0$ and $360 ^\circ$. The distribution of on-sky velocities and magnitudes of the injected synthetic objects are visualized in Figure \ref{fig:fakes}. 

Detectors that were successfully processed through the stage of PSF and WCS fitting are warped/resampled onto an identical pixel grid and coadded using the LSP \\\texttt{CompareWarpAssembleCoaddTask}, which compares the result of direct and PSF-matched warps to construct the coadd. This coadd acts as a template which is subtracted from the processed science images using the LSP \texttt{SubtractImageTask} based on the method of Alard-Lupton \citep{Alard_Lupton_1998}. The result are difference images in which ideally only non-static astronomical sources remain: galaxies and non-variable stars are removed whereas variable stars and moving objects remain. The image subtraction is typically not perfect, especially when the WCS solutions are not perfect, leaving behind poorly subtracted stars. The cores of bright stars are typically also not modeled well in the template and remain in the single epoch images after subtraction. These artifacts can be a source of contamination in detection catalogs derived from difference images. The difference images are the inputs to the algorithm as described.

\subsection{Source Detection} \label{sec:detection}

We perform source detection using the LSP \texttt{SourceDetectionTask}. This task uses a PSF detection model to determine the location of putative sources. An input image---pixel-level flux values and their estimated variance---are convolved with the PSF. A ratio of the convolved flux and variance is computed to form an image where each pixel represents the SNR of a putative detection at that pixel location. This SNR-image is thresholded at a provided SNR cutoff to form ``footprints'' or regions of pixels above-threshold.  Peaks in SNR are found in the footprints and reported as a putative source detection, including its pixel location and SNR. Properties of the source such as the brightness, shape, and location (at a sub-pixel level) can be further refined using model fitting. Detections are flagged and removed if affected by pixel-level defects, such being a bad pixel or near the edge of a CCD, or observational effects, such as being saturated or affected through crosstalk with nearby saturated pixels.

\begin{figure}[h]
    \centering
    \includegraphics[width=\linewidth]{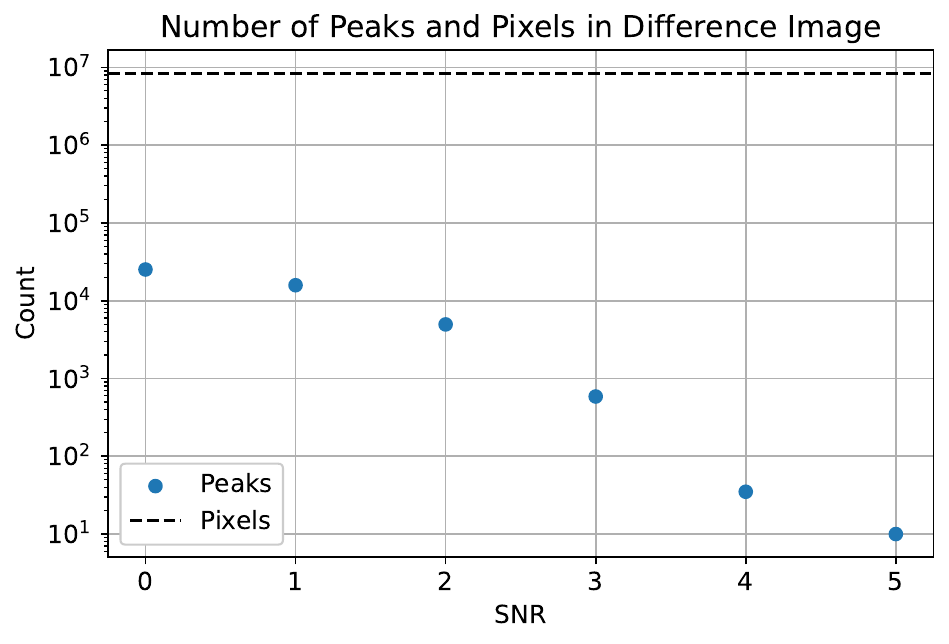}
    \caption{The number of peaks and pixels in a likelihood image derived from a $2048\times4096$ CCD difference image as a function of the SNR threshold. At all values of SNR, the number of detection peaks is smaller by factors of $10^2-10^6\times$.}
    \label{fig:peaks_and_pixels}
\end{figure}

\begin{figure*}[h!]
    \centering
    \includegraphics[width=\linewidth]{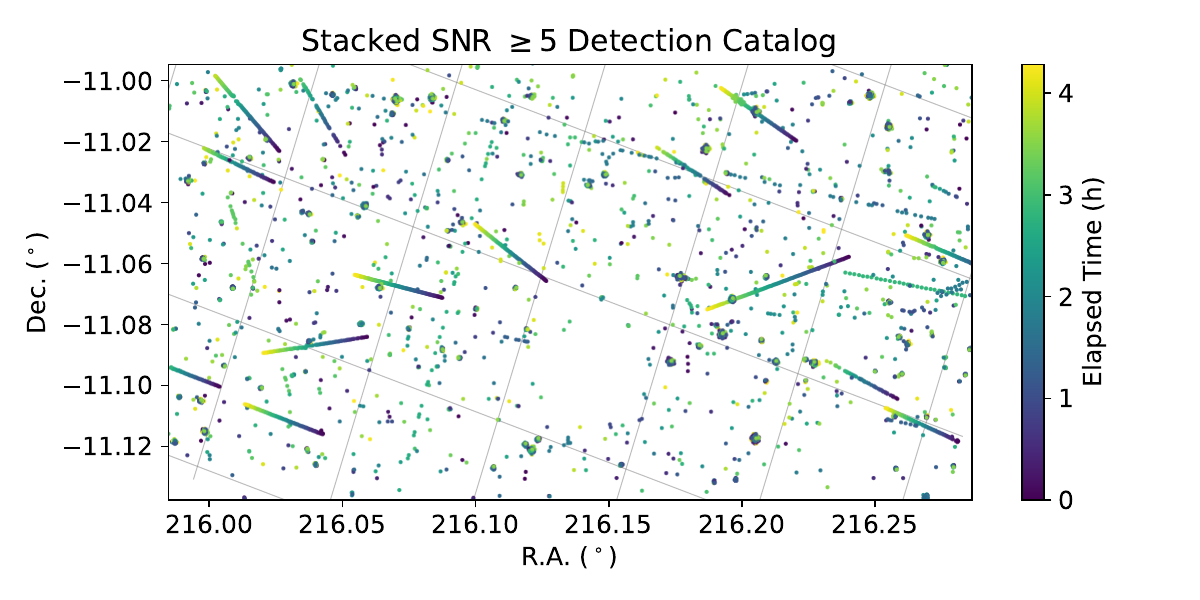}
    \caption{A visualization of the sky locations of a stacked detection catalog derived from a 4 hour sequence of DECam images with a grid of constant ecliptic latitude/longitude overlaid. Detections at a single epoch represent the on-sky location of a putative source on a difference image. Detections are accumulated across all 104 epochs and plotted jointly in this visualization. Moving objects appear as lines in this stacked visualization. Stationary variable sources as well as difference imaging artifacts appear as clusters and isolated points. Holes or voids in this visualization represent regions of the CCDs that were masked out due to the presence of bright stars and the resulting saturation of portions of the detector.}
    \label{fig:catalog}
\end{figure*}

Figure \ref{fig:peaks_and_pixels} visualizes the empirical number of peaks as a function of SNR threshold for a real difference image from DECam. The number of peaks is far fewer than the number of pixels in the image across all values of the SNR threshold. Detection peaks are associated with a sky-location RA $\alpha$ and Dec $\delta$ via a world coordinate system (WCS) fit to the exposure as well as an observation time $t$. For moving objects, the observation time is set to the midpoint of the exposure observation time, including the time to operate the shutter. Figure \ref{fig:catalog} visualizes the sky coordinates of a $\text{SNR}\geq5$ detection catalog derived from a sequence of $104$ DECam exposures limited to a single detector. Moving objects appear clearly as lines in this stacked visualization of the input detection catalogs. The goal of our algorithm is to recover these lines.

\subsection{Search} \label{sec:search}

We used our catalog shift-and-stack algorithm to search for TNOs and MBAs in the processed difference images. To search for TNOs, we used a velocity range of $0.003-0.04$ deg/day ($0.45 - 6$ arcsec/hour) and for MBAs a velocity range of $0.1 - 0.5$ deg/day. Test trajectories were generated using these velocity ranges and covering all on-sky angles. These rates cover the span of angular velocities of synthetic sources injected into the images. We search for objects in each DECam detector in the single long-stare pointing. Of the 62 detectors in the DECam focal plane, we exclude detector 2 since it is masked entirely in the data reduction process and detector 61 since it has data quality issues. We searched for moving objects in detection catalogs constructed using different SNR thresholds. The value of $\Delta x$ was varied among the MBA and TNO searches for each SNR threshold used. Table \ref{tab:search_params} enumerates the values of $\Delta x$ used for each search. $\Delta x$ was limited to $10$ and the SNR of the detection catalog to $3$ for the MBA search due the excessive computational cost involved. For each search performed, the first $1000$ clusters with the highest number of detections are output. Each cluster is passed through the MCD regression step, resulting in a refined estimate of the candidate object's trajectory. For each refined trajectory guess, detections are gathered from the per-epoch detection catalogs that are within a positional distance threshold of $\Delta x$.

\begin{table}[]
    \centering
    \begin{tabular}{c|c|c}
         & \multicolumn{2}{c}{$\Delta x$ ($a$)} \\
        \hline
        $\nu$ &  TNO & MBA \\
        \hline 
        1 & 1 & - \\
        2 & 1 & - \\
        3 & 3 & 10 \\
        4 & 10 & 10 \\
        5 & 10 & 10
    \end{tabular}
    \caption{Values of $\Delta x$ in units of the PSF width $a$ for each search performed. A value of - indicates the search was not performed for the combination of $\nu$ and population.}
    \label{tab:search_params}
\end{table}

\subsection{Performance} \label{sec:performance}

We measure the performance of our algorithm by quantifying its ability to recover injected synthetic objects in the searches performed and by measuring the runtime and memory usage of the searches.  

\subsubsection{Depth and Completeness}

We measure the ability of our algorithm to recover injected synthetic objects by quantifying the completeness of the search. We measure completeness as a function of magnitude using a parametrized model as in \cite{Bernardinelli_2024}
\begin{align} \label{eqn:completeness}
    p(m| c, k, m_{50}) &= \frac{c}{1 + \exp{(k(m-m_{50}))}}
\end{align}
where $m_{50}$ is the magnitude at 50\% completeness, $c$ is the fraction of objects detected when $m \ll m_{50}$, and $k$ is a steepness parameter that encodes how rapid the drop-off in completeness is at $m_{50}$. The parameter values are estimated by maximizing the likelihood
\begin{align}
    \mathcal{L} &= \prod_{i \in \mathcal{D}} p(m_i | c, k, m_{50}) \prod_{i\in\mathcal{D}^c} \left(1 - p(m_i | c, k, m_{50})  \right)
\end{align}
where $\mathcal{D}$ is the set of synthetic objects recovered, $\mathcal{D}^c$ is the set of synthetic objects not recovered,  and $m_i$ is the magnitude associated with a single synthetic object.

Search candidates are matched to synthetic objects by comparing their trajectories. Trajectories are matched by comparing the differences in the positions predicted by the trajectories at each epoch. Two trajectories are considered matched if their predicted positions differ by less than $1$ arcsec in 50\% or more of the epochs. A synthetic object is considered recovered if its trajectory is successfully matched to a candidate trajectory from our algorithm.

A single-epoch $\text{SNR} \sim5$ limiting magnitude is estimated by matching the catalog of synthetic objects to detected sources in a $\text{SNR}\geq 5$ detection catalog using a $1$ arcsec matching radius. For each synthetic object injected, the fraction of times it appears in the $\text{SNR}\geq 5$ detection catalog is measured. This fraction follows the same statistics as the survey completeness, allowing us to model it as a function of magnitude using Eq.~\ref{eqn:completeness}. This produces a single-epoch $\text{SNR}\sim 5$ limiting magnitude of $m_{50} = 23.7$. A theoretical upper limit on the $m_{50}$ depth achievable through coaddition can be found by considering the relative flux of a single-epoch detection at SNR $\nu$ and a coadded detection at the same SNR $\nu$: 
\begin{align}
    \nu &= \frac{f}{\sigma} = \frac{f_{\mathrm{coadd}}}{\sigma_{\mathrm{coadd}}} \\
    &\rightarrow \frac{f}{f_{\mathrm{coadd}}} = \frac{\sigma}{\sigma_{\mathrm{coadd}}}~.
\end{align}
If we assume constant background noise and image quality, the uncertainty in the coadd scales as
\begin{align}
    \sigma_{\mathrm{coadd}} &= \sqrt{\frac{\sigma^2}{N_t}} \\
    \sqrt{N_{t}} \sigma_{\mathrm{coadd}} &= \sigma
\end{align}
which lets us compute the relative single-epoch and coadded flux as
\begin{align}
    \frac{f}{f_{\mathrm{coadd}}} &= \frac{\sqrt{N_{t}}\sigma_{\mathrm{coadd}}}{\sigma_{\mathrm{coadd}}} = \sqrt{N_{t}}
\end{align}
and is equivalent to a magnitude difference of
\begin{align}
    m_{50}^{\mathrm{coadd}} - m_{50} &= \left(-\frac{5}{2} \log_{10}(f_{\mathrm{coadd}}) + z_p\right) - \left(-\frac{5}{2} \log_{10}(f) + z_p\right) \\
    &= \frac{5}{2} \log_{10}\left(\frac{f}{f_{\mathrm{coadd}}}\right) \\
    &= \frac{5}{2} \log_{10}(\sqrt{N_{t}})
\end{align}
where $z_p$ is a shared photometric zero-point. Given $N_{t} = 104$, we find that $m_{50}^{\mathrm{coadd}} = m_{50} + \frac{5}{2}\log_{10}(\sqrt{N_t}) = 26.2$.

\begin{figure*}[h]
    \centering
    \includegraphics[width=\linewidth]{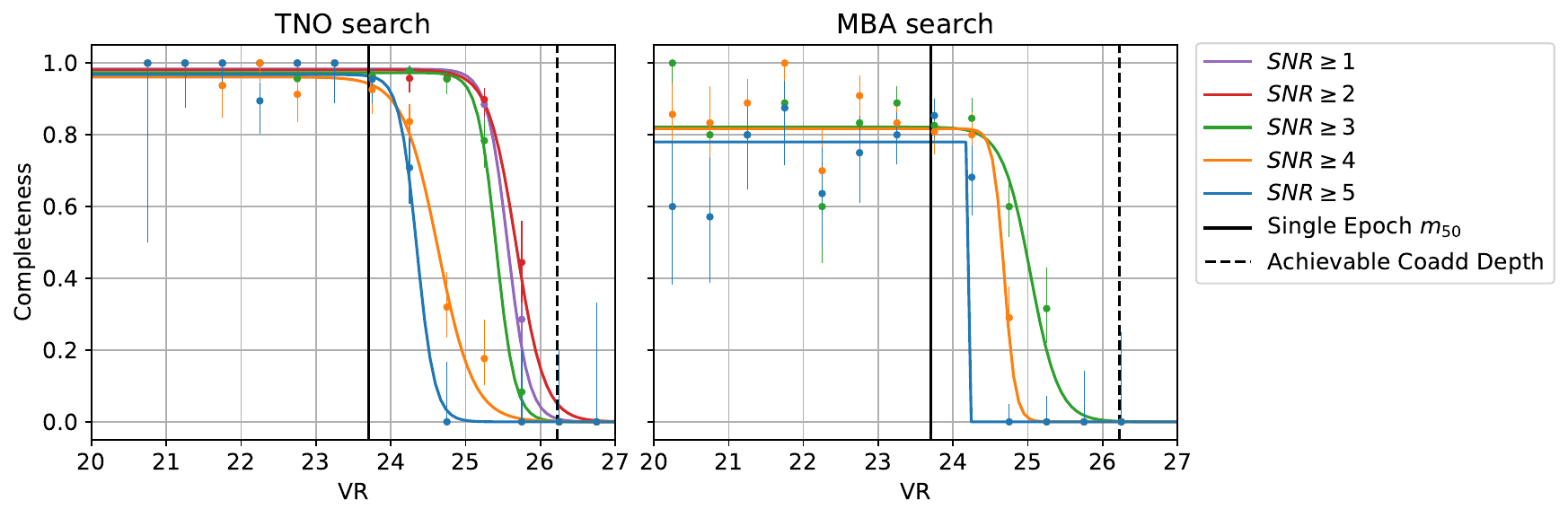}
    \caption{The fraction of synthetic objects recovered as a function of their magnitude for the TNO and MBA searches. Dots represent the detection fraction in bins of size $0.5$ mag and error bars represent the asymmetric uncertainty in the estimated fraction using the Wilson score interval. Two reference magnitudes are visualized as black solid and dashed vertical lines. The solid black line visualizes the estimated single-epoch $m_{50} = 23.7$. The dashed black line represents the theoretical achievable depth by coadding all of the images and is equal to $m_{50}^{\mathrm{coadd}} = 26.2$.}
    \label{fig:completeness_tno_mba}
\end{figure*}

Figure \ref{fig:completeness_tno_mba} visualizes the completeness of the TNO and MBA searches performed after matching the 1000 candidate trajectories per-search to synthetic object trajectories. The solid black line in this figure represents the estimated single-epoch $\text{SNR}\geq 5$ limiting magnitude of $m_{50} = 23.7$ while the dashed black line represent the theoretical achievable depth of an optimal coadd at $m_{50}^{\mathrm{coadd}} = 26.2$.

\begin{figure*}[h]
    \centering
    \includegraphics[width=\linewidth]{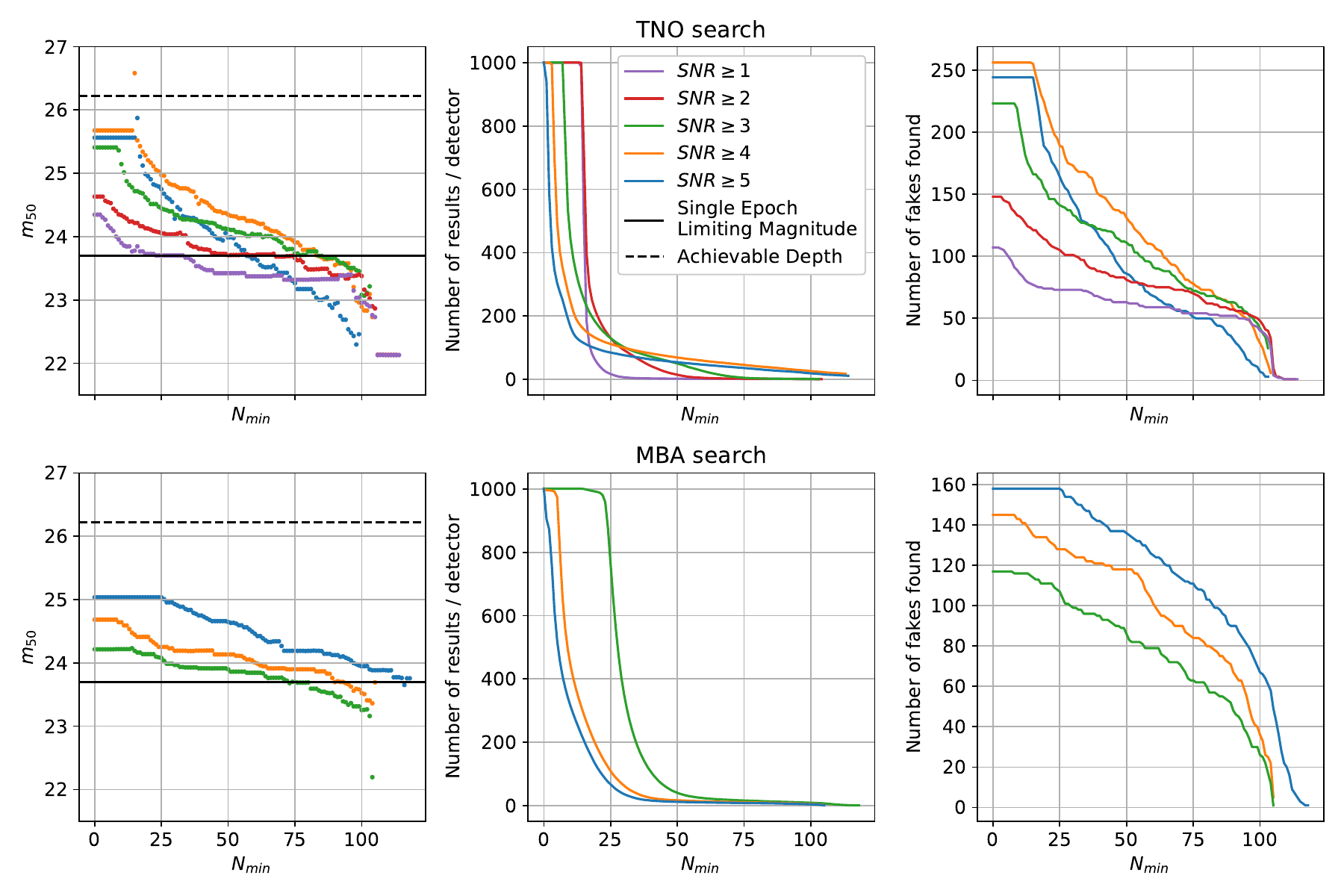}
    \caption{The $m_{50}$ depth from fitting of the completeness (left), the number of candidate detections produced per detector (middle), and the number of synthetic objects (fakes) recovered (right) for the TNO and MBA searches as a function of the $N_{\mathrm{min}}$ parameter. The solid black line visualizes the estimated single-epoch $m_{50} = 23.7$. The dashed black line represents the theoretical achievable depth by coadding all of the images and is equal to $m_{50}^{\mathrm{coadd}} = 26.2$. Decreasing the value of $N_{\mathrm{min}}$ increases $m_{50}$ depth and the number of synthetic objects recovered, at the cost of vastly increasing the number of results produced. Most of the results produced at low values of $N_{\mathrm{min}}$ are false-positive candidates.}
    \label{fig:m50_Nmin}
\end{figure*}

\begin{figure}[h]
    \centering
    \includegraphics[width=\linewidth]{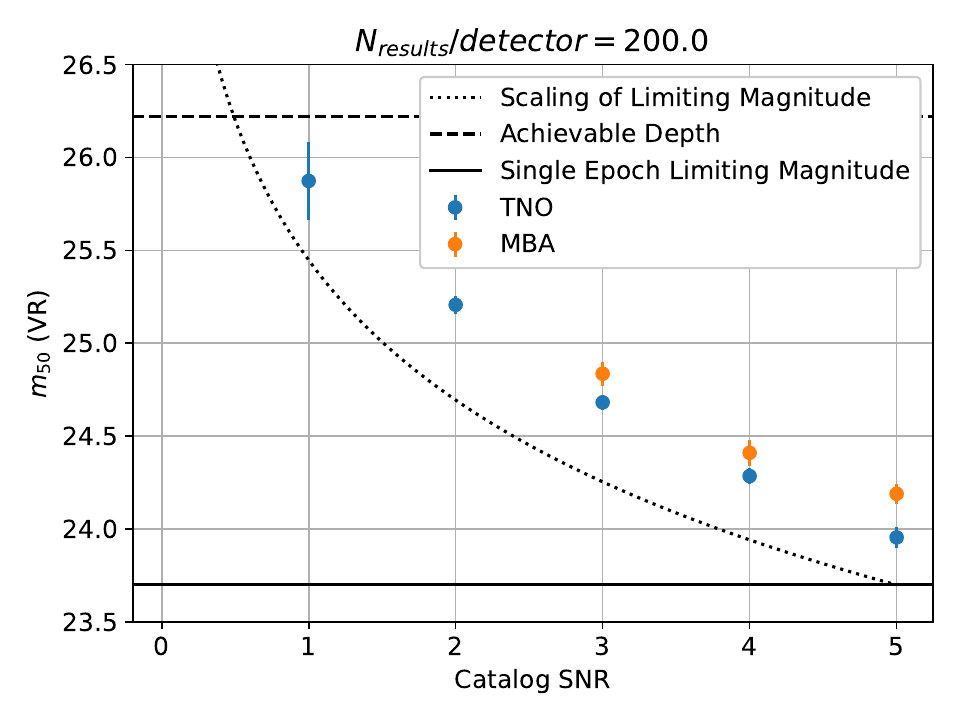}
    \caption{The $m_{50}$ depth achieved for the TNO (blue) and MBA (orange) searches as a function of the SNR of the input catalog choosing $N_{\mathrm{min}}$ such that the number of results per detector is fixed at $200$. The solid black line represents a single-epoch $\text{SNR}\sim5$ limiting magnitude of $m_{50} = 23.7$ while the dashed black line represents the theoretical optimal achievable coadded depth of $m_{50}^{\mathrm{coadd}} = 26.2$. The dotted black line visualizes the expected scaling of achieved depth with SNR, extrapolated from the single-epoch limiting magnitude.}
    \label{fig:m50_SNR}
\end{figure}

The completeness varies as a function of the parameter $N_{\mathrm{min}}$. Greater $m_{50}$ depth can be achieved by decreasing the value of $N_{\mathrm{min}}$ at the cost of increasing the number of false results included. This is a choice that the operator of this algorithm must make, and will depend on the operator's desired false positive rate. This trade-off is explored by choosing an $N_{\mathrm{min}}$ cutoff and removing candidate trajectories from the set of $1000$ candidates per search. For each value of $N_{\mathrm{min}}$, the $m_{50}$ depth, the number of candidate trajectories, and the number of fakes found are evaluated. These values as a function of $N_{\mathrm{min}}$ are visualized in Fig.~\ref{fig:m50_Nmin}. We find that decreasing the value of $N_{\mathrm{min}}$ increases $m_{50}$ depth and the number of synthetic objects recovered, as expected. However, this comes at the cost of vastly increasing the number of results produced. Most of the numerous results produced at low values of $N_{\mathrm{min}}$ are likely false-positive candidates due to stacks of noise detections. The scaling of achievable $m_{50}$ depth with the SNR of the input catalog is visualized in Fig.~\ref{fig:m50_SNR} and enumerated in Tab.~\ref{tab:m50_SNR} by choosing a value of $N_{\mathrm{min}}$ for each search that corresponds to $200$ results per CCD. 

\begin{table}[]
    \centering
    \begin{tabular}{c|c|c}
        SNR & TNO $m_{50}$ & MBA $m_{50}$ \\ \hline
        5   & $23.96 \pm 0.06$ & $24.19 \pm 0.05$ \\ \hline
        4   & $24.29 \pm 0.05$ & $24.41 \pm 0.07$ \\ \hline
        3   & $24.68 \pm 0.04$ & $24.84 \pm 0.06$ \\ \hline
        2   & $25.21 \pm 0.05$ & - \\ \hline
        1   & $25.87 \pm 0.21$ & -
    \end{tabular}
    \caption{The $m_{50}$ depth achieved in the TNO and MBA searches of a catalog of the given SNR. A value of - indicates the search was not performed.}
    \label{tab:m50_SNR}
\end{table}

\subsection{Computational Cost}

\begin{figure}[h]
    \centering
    \includegraphics[width=\linewidth]{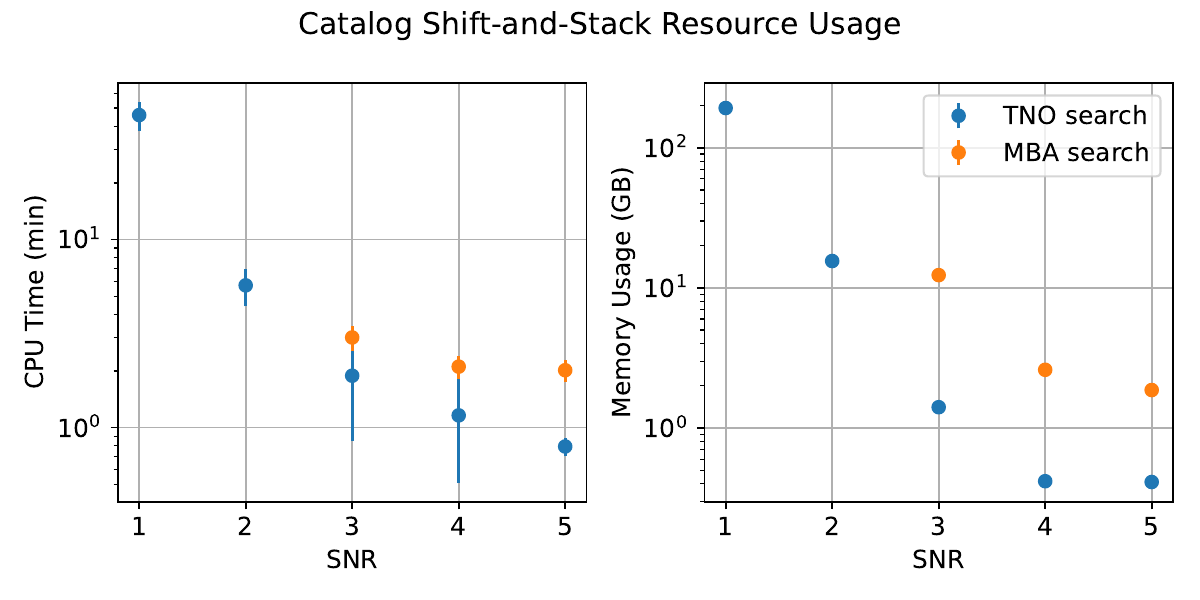}
    \caption{The median wall-clock runtime and memory usage of our catalog shift-and-stack algorithm for the TNO (blue) and MBA (orange) searches across the $60$ DECam detectors searched. Error bars represent the standard deviation of the runtime and memory usage values measured.}
    \label{fig:cost}
\end{figure}

The computation cost of our algorithm scales from negligible to vast depending on the range of velocity trial trajectories, the SNR of the detection catalog, and the value of $\Delta x$. Figure ~\ref{fig:cost} visualizes the median wall-clock runtime and memory usage among the searches performed for TNOs and MBAs. Wall-clock time includes the time just for the algorithm operations, and excludes the time for source detection on the input images. The search runtime is $\sim 1$ min ($0.017$ core-hours) for SNR 3-5 detection catalogs using the values of $\Delta x$ in Table~\ref{tab:search_params} whereas the memory usage is $\sim1-10$ GB for the same runs. The computational cost balloons for the TNO searches at SNR 1 and 2, where $\Delta x = 1$ in which case the memory usage is $\sim 10-100$ GB and search wall-clock times can range from $\sim 10-40$ minutes ($0.17-0.67$ core-hours). 

\begin{figure}
    \centering
    \includegraphics[width=\linewidth]{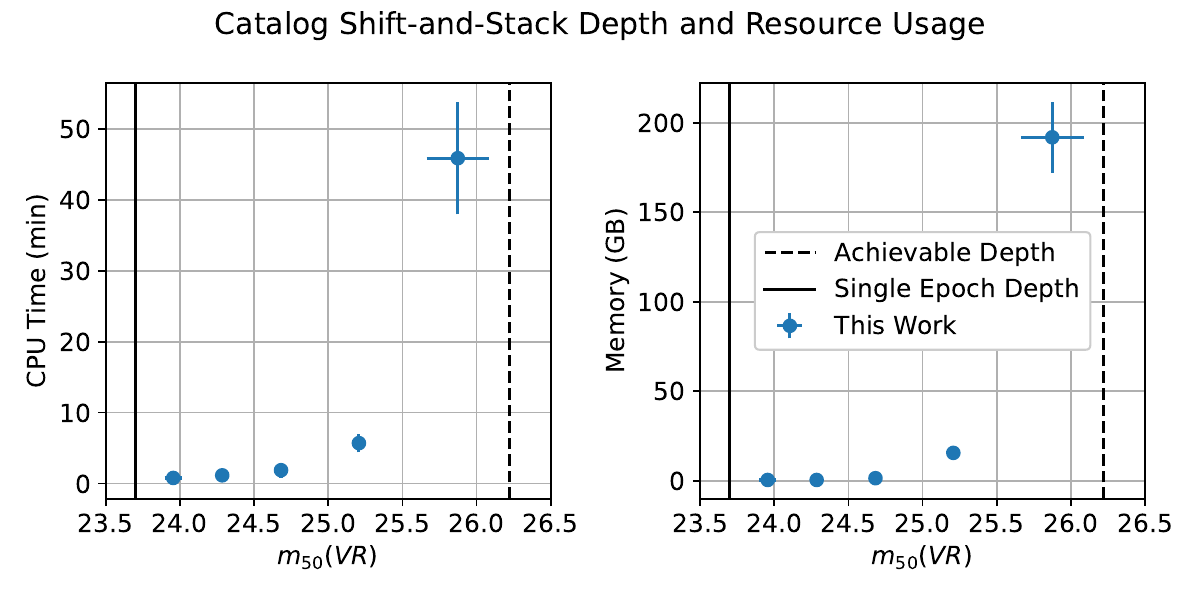}
    \caption{The CPU-time and memory usage of our algorithm as a function of the $m_{50}$ depth achieved. The black solid line visualizes the single-epoch $m_{50} = 23.7$ while the dashed black line visualizes the theoretical maximum $m_{50}^{\mathrm{coadd}} = 26.2$ that could be achieved through optimal coaddition.}
    \label{fig:scaling_1}
\end{figure}

\begin{figure}
    \centering
    \includegraphics[width=\linewidth]{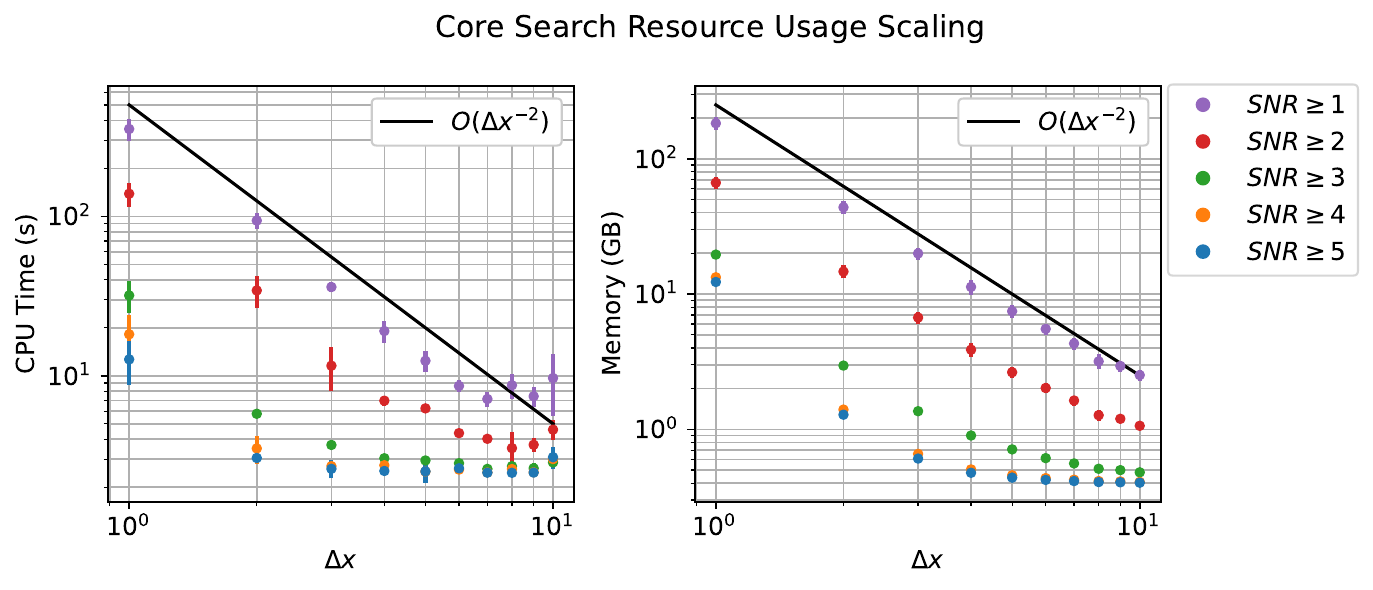}
    \caption{The CPU-time and memory usage of the core search component of our algorithm as a function of the $\Delta x$ value and the SNR of the catalog searched. The black lines represent the predicted scaling of CPU-time and memory with $\Delta x^{-2}$.}
    \label{fig:scaling_dx_1}
\end{figure}

In Fig.~\ref{fig:scaling_1}, we explore the compute cost and $m_{50}$ depth trade off of our algorithm by measuring the CPU-time and memory required to perform the TNO searches and comparing that to the $m_{50}$ depth achieved, assuming 200 results per-detector are reported.  We report only the amount of CPU time spent in performing the core-search component of our algorithm, excluding image loading, source detection, and trajectory refinement. The memory usage reported is the maximum amount of memory used during this procedure. The scaling of CPU time and memory usage as a function of $\Delta x$ is also visualized in Fig.~\ref{fig:scaling_dx_1}, verifying the scaling relationships predicted by Eq.~\ref{eqn:scaling}.

\section{Comparison With Image-Based Shift-and-Stack} \label{sec:comparison}

We attempt to compare the performance of our algorithm, a catalog-based shift-and-stack, to an image-based shift-and-stack algorithm. We selected the \texttt{KBMOD} software framework to perform our comparison \citep{Whidden_2019, jeremy_kubica_2023_7849953}.\footnote{\texttt{KBMOD} is available online at \url{https://github.com/dirac-institute/kbmod/}} \texttt{KBMOD} is an open-source software framework for performing shift-and-stack searches accelerated with Graphics Processing Units (GPUs). \texttt{KBMOD} convolves a PSF-model with the provided images to construct images with pixel-values representing detection likelihood at that pixel location, identical to the source-detection method outlined in Section \ref{sec:detection}. The input images are warped into a common pixel reference frame using the per-image WCS. The pixels from each ``likelihood image'' are then shifted according to linear motion (in pixels per unit time) with respect to each pixel in the first image of the stack. For each shift performed, a sum is computed across the images. This summed value corresponds to a coadded detection likelihood at a given pixel location and for a given linear trajectory.

We used \texttt{KBMOD} to search for TNOs in the same set of images that our algorithm was run on. We used a similar range of on-sky velocity rates as in Section \ref{sec:search} but covering 180$^\circ$ in angle around the ecliptic as opposed to the 360$^\circ$ angular range searched with our algorithm. Velocity rates are sampled densely enough so the velocity guesses diverge by no more than 1 arcsecond over the timespan of the data. We output results with a coadded likelihood of 10 or more. 

\subsection{Depth and Completeness}

\begin{figure}[h]
    \centering
    \includegraphics[width=\linewidth]{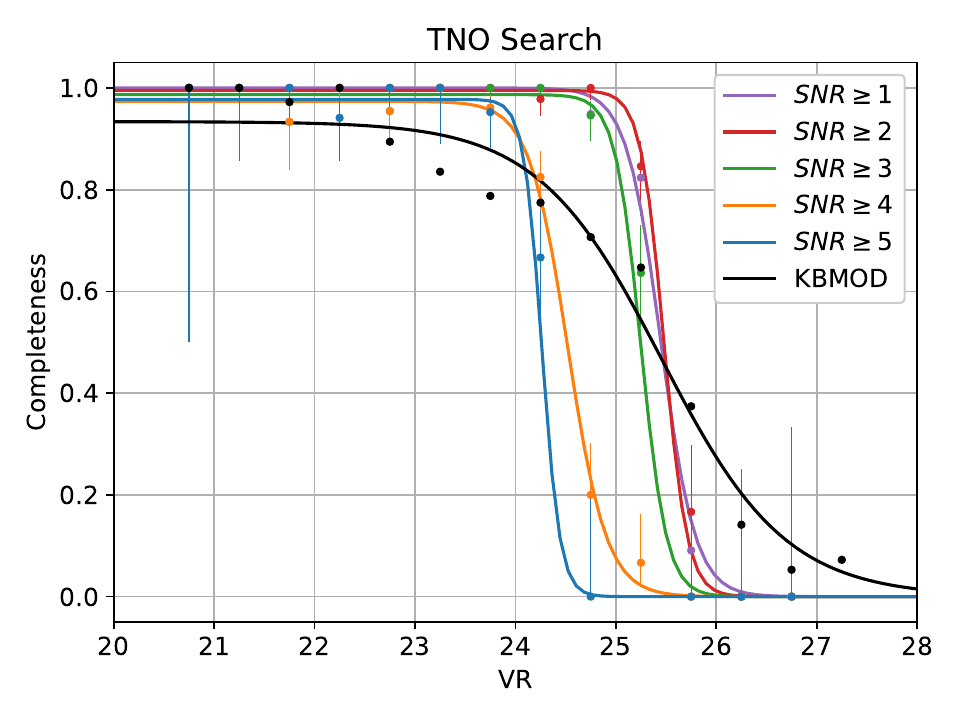}
    \caption{The fraction of synthetic TNOs recovered as a function of their magnitude using \texttt{KBMOD} and our algorithm at SNR $\geq 5, 4, 3, 2, 1$.}
    \label{fig:tno_search}
\end{figure}

Results from the \texttt{KBMOD} search were matched to injected synthetic objects using the same method and criteria as outlined in Section \ref{sec:performance}. From this, a completeness curve is fit using Eq.~\ref{eqn:completeness} to characterize the performance of the search. This curve is visualized along with the measured completeness of the TNO search using our algorithm in Fig.~\ref{fig:tno_search}. 

The \texttt{KBMOD} search achieves a depth of $m_{50} = 25.47 \pm 0.01$. The $\text{SNR}\geq 5, 4, 3, 2$ searches we performed produce an $m_{50}$ depth that is smaller than the \texttt{KBMOD} search. Our searches at $\text{SNR}\geq1$ surpass the $m_{50}$ depth of the \texttt{KBMOD} search, achieving a depth of $m_{50} = 25.87 \pm 0.21$. These numbers provide only rough depth comparison between the catalog and image-based shift-and-stack approach, as we make no attempt to compare or achieve parity between the two implementations in terms of false positive and true positive rates, which are ultimately the factors that matter when deriving scientific insight from these searches. However, we can conclude that our method of stacking detection catalogs can achieve comparable depths as an image-based approach in a practical sense.

\subsection{Accuracy of Trajectories}

We additionally compare the accuracy of the recovered synthetic object trajectories between \texttt{KBMOD} and our algorithm. The accuracy is quantified for each recovered trajectory by calculating the great-circle distance between the ground truth (injected) sky location and the predicted sky location using the recovered trajectory at each injection epoch. The average separation is computed by dividing the total separation across epochs by the number of injection epochs. The metric is accumulated across all synthetic objects recovered by \texttt{KBMOD} and each of the SNR $\geq 5, 4, 3, 2, 1$ searches performed with our algorithm. 

\begin{figure}
    \centering
    \includegraphics[width=\linewidth]{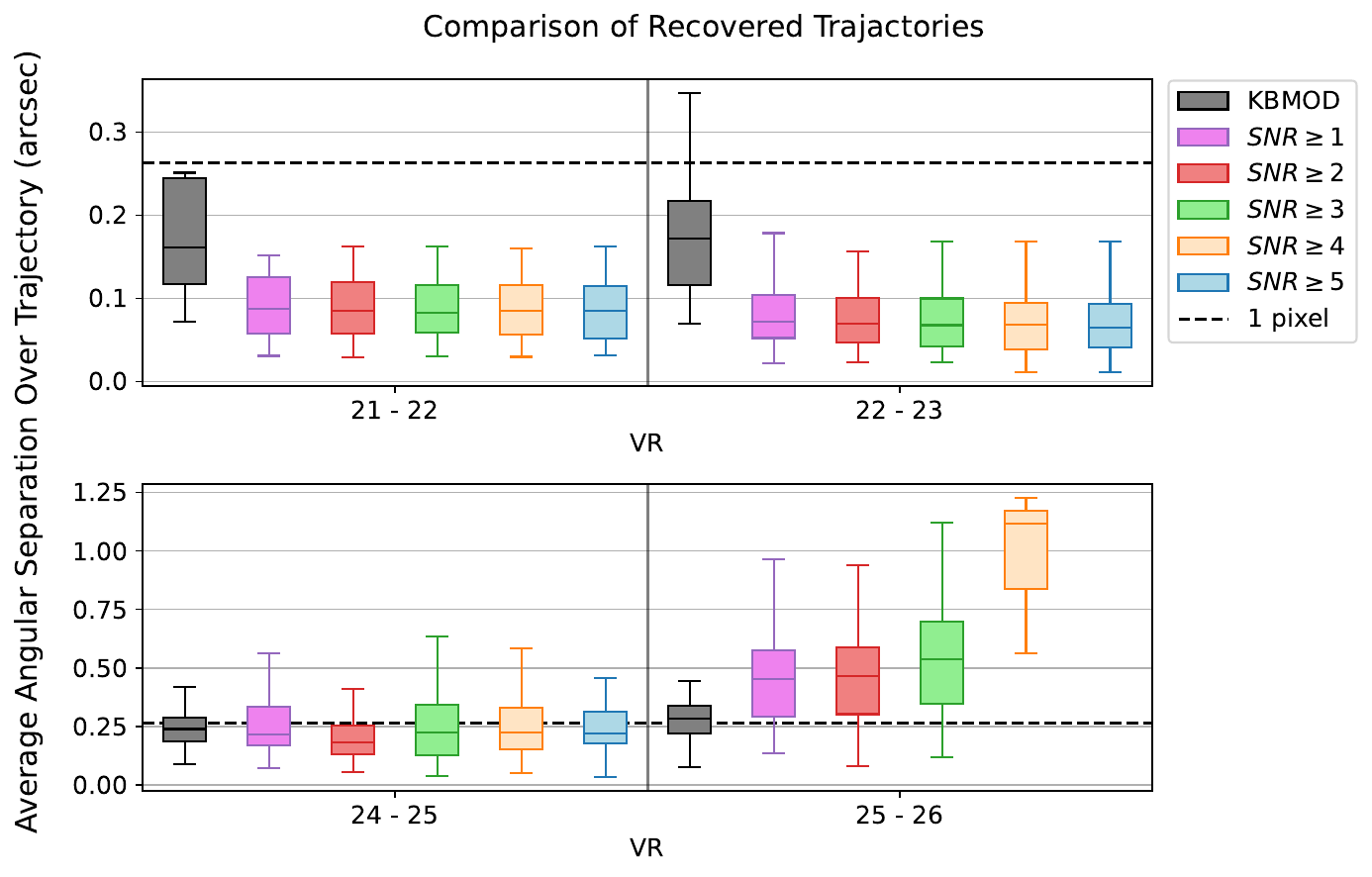}
    \caption{Distributions of average separation between the ground truth and recovered trajectories of synthetic objects found via the image-based shift-and-stack (\texttt{KBMOD}) and catalog-based shift-and-stack (this work) over SNR $= 1- 5$. Separation is computed as the great-circle distance between the two trajectories at a given point in time, and an average separation is computed over the 4-hour observing period. Distributions are represented as box-and-whisker plots which visualize (excluding outliers) the minimum, maximum, and 25th, 50th (median), and 75th percentiles of the distribution. A black dashed line visualizes an angular separation that corresponds to a fiducial pixel-scale for DECam $p_s = 0.263~\text{arcsec}/\text{pixel}$.}
    \label{fig:average_separation}
\end{figure}

Figure \ref{fig:average_separation} visualizes the distribution of average separation between the ground truth and recovered trajectories as a function of the injected object magnitude. A fiducial pixel scale for DECam $p_s = 0.263$ arcsec/pixel is visualized as well. We find that both algorithms achieve sub-pixel accuracy of the recovered trajectories for synthetic objects with magnitudes $21 - 25$, however the average separation between the \texttt{KBMOD} trajectories and the ground-truth is generally larger by a factor of $1.5-2\times$ that the trajectories from our algorithm for objects with magnitudes $21 - 23$. At the faint end, for objects with magnitudes $25-26$, \texttt{KBMOD} retains its accuracy of $\sim1$ pixel, while the accuracy of the trajectories obtained from our algorithm diverges with SNR. This is expected, as these faint objects are unlikely to appear at all in a high-SNR (e.g.~SNR$\geq5$) catalog, and thus the trajectory estimates derived from a fit of their positions will be subject to noise due to the lack of available data. Additionally, even if the number of data points is increased in a low-SNR (e.g.~SNR$\geq 1$) catalog, the cataloged positions of these objects will be uncertain (scaling inversely with SNR per Eq.~\ref{eqn:pos_uncertainty}), which propagates into the trajectory estimated from a fit to these positions.

\subsection{Computational Cost}

\begin{figure}
    \centering
    \includegraphics[width=\linewidth]{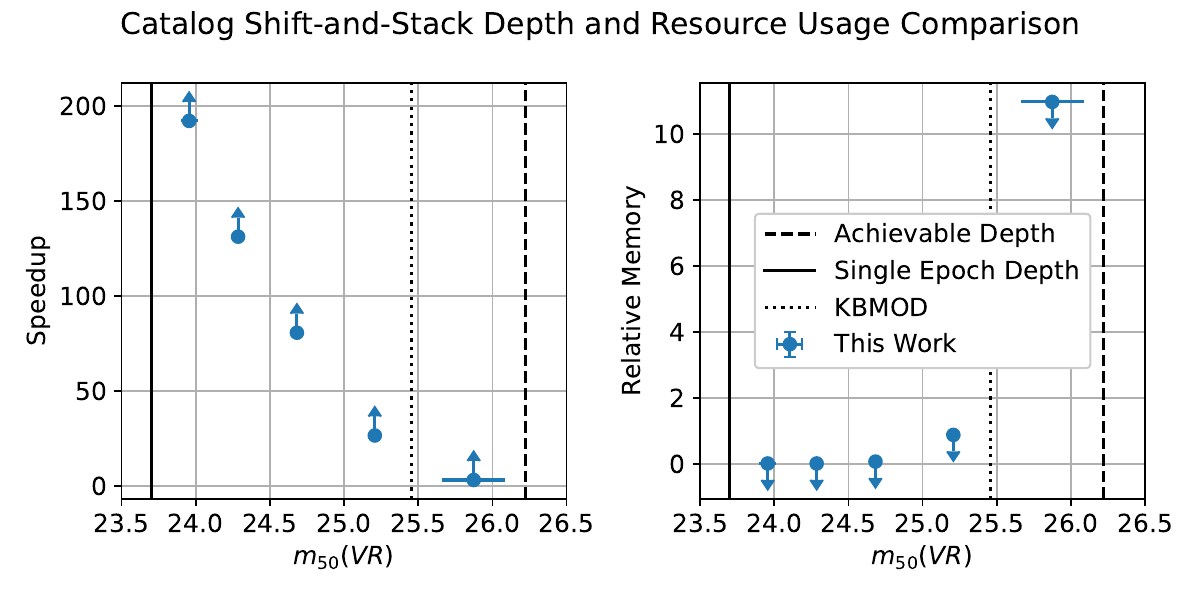}
    \caption{The relative total CPU-time (speedup) and memory usage of our algorithm when compared to an image-based shift-and-stack search. The black solid line visualizes the single-epoch $m_{50} = 23.7$ while the dashed black line visualizes the theoretical maximum $m_{50}^{\mathrm{coadd}} = 26.2$ that could be achieved through optimal coaddition. The dotted black line visualizes the $m_{50}^{\mathrm{KBMOD}} = 25.47$ depth achieved in an image-based shift-and-stack approach, utilizing the \texttt{KBMOD} software.}
    \label{fig:scaling_2}
\end{figure}

\begin{figure}
    \centering
    \includegraphics[width=\linewidth]{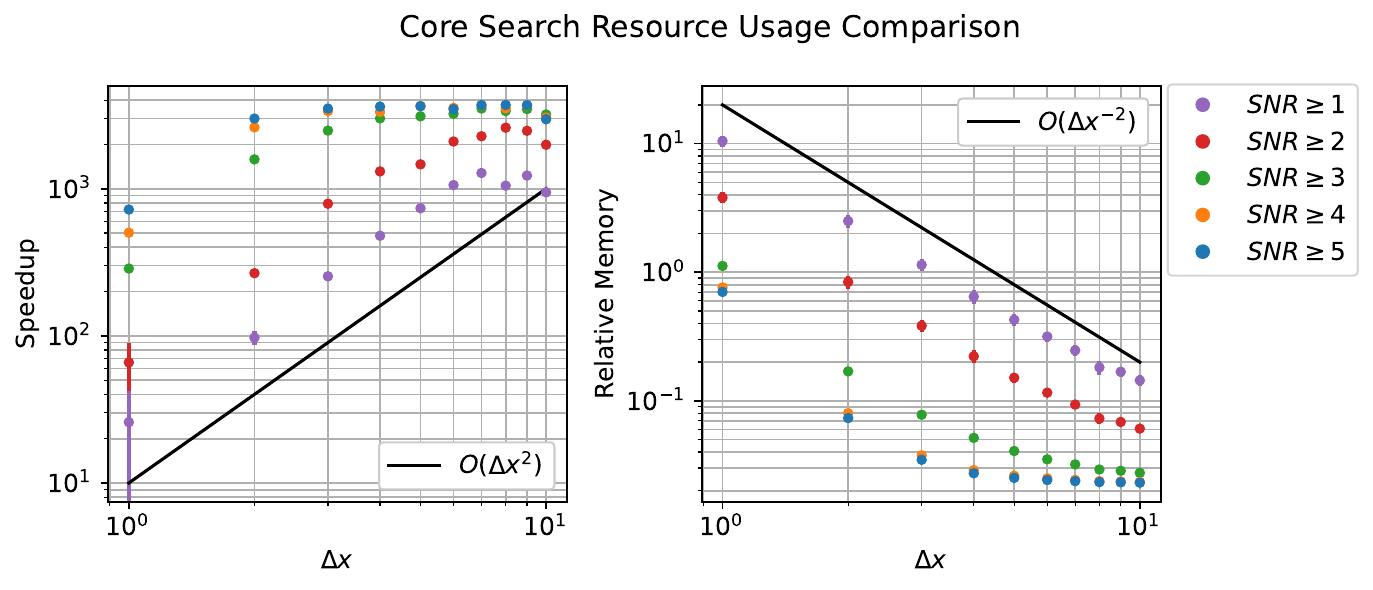}
    \caption{The relative core-search CPU-time (speedup) and memory usage of our algorithm compared to an implementation of an image-based shift-and-stack search. The black lines represent the predicted scaling of speedup and memory with $\Delta x^2$ and $\Delta x^{-2}$ respectively.}
    \label{fig:scaling_dx_2}
\end{figure}

The \texttt{KBMOD} software is GPU-accelerated and provides no CPU-only implementation of its search, which makes it challenging to perform a compute-time comparison between these two methods. In order to make this comparison explicitly, we produced a CPU-implementation of the approach outlined in \cite{Whidden_2019}. Our implementation includes convolution of the inputs images, shifting pixels according to provided input trajectories, summing pixels to compute a coadded detection SNR, and identifying coadded pixels that are above a provided threshold, corresponding to a putative detection for the given trajectory. In this implementation, we perform and measure the CPU-time required for the pixel shifts, additions, divisions, and thresholding, excluding the actual reporting or filtering of results. This allows us to make a direct comparison of the compute-time (i.e.~number of operations) required for the core search components of the catalog and image-based shift-and-stack approaches. We performed a single search using the same velocity and angle ranges used in our TNO-search with a trajectory divergence of no more than 1 arcsecond over the timespan of the data. The CPU-time required for the search was 2.54 hours and required 17.5 GB of memory.  While the CPU-time measurement is likely faithful to the expected runtime of the image-based shift-and-stack, the memory usage can likely be reduced in a more optimized implementation. These values are lower-limits on the total compute-time and memory usage of the image-based shift-and-stack, since we do not find and report detection results in our CPU-only implementation. These measurements are used to compute the relative CPU-time (speedup) and relative memory usage between our detection catalog stacking algorithm and an image-based shift-and-stack approach, producing lower and upper limits on these values respectively. Figure \ref{fig:scaling_2} visualizes the total compute-time speedup and relative memory usage of our algorithm relative to an image-based shift-and-stack search as a function of $m_{50}$ depth, while Fig.~\ref{fig:scaling_dx_2} visualizes the speedup and relative memory usage when comparing the core-search components of the catalog and image-based shift-and-stack approach. Examining Fig.~\ref{fig:scaling_dx_2}, we find that the core-search component of our algorithm attains speedups of $\sim10-10^3\times$ across values of $\Delta x$ from $1 -10$ and SNR $1-5$. 

Values of the absolute and relative total CPU-time, memory usage, and $m_{50}$ depth achieved by our algorithm and the image-based shift-and-stack approach are reported in Table~\ref{tab:speedup}. We find that total compute-time speedups range from $3.9\times$ for the SNR $\geq 1$ search to $203.3\times$ for the SNR $\geq 5$ search, providing speedup over all achievable $m_{50}$ depths. Our algorithm uses $10.9\times$ as much memory for the SNR $\geq 1$ search and $50 \times$ less memory for the SNR $\geq 5$ search, realizing memory savings for most searches except the SNR $\geq 1$ search, which has a likely unmanageable memory footprint.  Taking the SNR $\geq 2$ search as a comparison point, our algorithm achieves a speedup of $30.8\times$ with $88\%$ of the memory footprint of an image-based shift-and-stack approach, while sacrificing $0.25$ mag in $m_{50}$ depth. 

\begin{table*}[]
    \centering
    \begin{tabular}{c|c|c|c|c|c|c}
        SNR & $m_{50}$ & CPU (min) & Memory (GB) & Speedup & Relative Memory & $\Delta m_{50}$ \\ \hline
1 & $25.87 \pm 0.21$ & 39.5 & 190.8 & $\geq 3.9 \times$ & $\leq 10.92 \times$ & $0.42 \pm 0.21$ \\ \hline
2 & $25.21 \pm 0.05$ & 5.0 & 15.4 & $\geq 30.8 \times$ & $\leq 0.88 \times$ & $-0.25 \pm 0.05$ \\ \hline
3 & $24.68 \pm 0.04$ & 1.8 & 1.4 & $\geq 85.9 \times$ & $\leq 0.08 \times$ & $-0.77 \pm 0.04$ \\ \hline
4 & $24.29 \pm 0.05$ & 1.1 & 0.4 & $\geq 138.6 \times$ & $\leq 0.02 \times$ & $-1.17 \pm 0.05$ \\ \hline
5 & $23.96 \pm 0.06$ & 0.8 & 0.4 & $\geq 203.3 \times$ & $\leq 0.02 \times$ & $-1.50 \pm 0.06$
\end{tabular}
    \caption{Total compute-time, memory usage, and $m_{50}$ depth achieved when searching for TNOs. Speedup, relative memory usage, and difference in $m_{50}$ depth in comparison to the core-search routine of an image-based shift-and-stack approach are provided, using 2.54 core-hours, 17.5 GB of memory, and $m_{50} = 25.47 \pm 0.01$ as fiducial values.}
    \label{tab:speedup}
\end{table*}

\section{Conclusions}

We have presented an efficient shift-and-stack algorithm capable of finding both bright and faint moving objects in detection catalogs derived from CCD images. The algorithm finds objects in both pure and noisy catalogs derived from high- and low-SNR detections. The algorithm is efficient in the sense that it reduces the number of stacks performed relative to an image-based shift-and-stack algorithm.

We have validated the performance of the algorithm by using it to recover synthetic moving objects implanted in difference images representing the Main Belt Asteroid and Trans-Neptunian Object populations. The algorithm is found to be computationally efficient and effective at recovering faint moving objects below the noise floor of an individual image. We provide a method to choose the parameters of the algorithm that determine the depth it can achieve as well as its false positive rate. We show how depth achieved scales with the number of candidate objects discovered, and the required resource usage to attain a given depth. We provide a comparison of resource usage and depth obtained between our method and an image-based shift-and-stack approach. We find that our method can achieve wall-clock runtime speedups of $\sim30\times$ with $88\%$ of the memory usage while sacrificing $0.25$ mag in $m_{50}$ depth.

The algorithm does not reach the theoretical limit expected from ideal coaddition. One possible reason for this is due to our limited method of results filtering. In this study, we analyze depth based on the first 1000 candidates produced in a search. Applying a more sophisticated filtering method on the produced results may allow fainter objects of lower significance to be found while keeping the number of results manageable. Additionally, it is possible that this is due to the metric used for accumulating signal: the number of detections across epochs in an aperture of variable size. A different metric, perhaps the detection likelihood could be accumulated instead, mimicking the approach of \cite{Whidden_2019}. The likelihood naturally weights signal detections higher than noise detections, whereas the indicator on detected vs non-detected provides equal weight to these outcomes. In this scheme, a likelihood ratio test could be used to reason about the relative odds that a candidate moving object is real or the result of noise detections, following more closely the detection catalog coaddition framework outlined in \cite{Budavári_2017}, which was able to clearly distinguish real faint objects from noise detections at very low detection significance. A benefit of our equal-weighting scheme is that it potentially makes our shift-and-stack method less sensitive to contamination by difference image artifacts and stationary sources. Exploring the cost-benefit trade-off of using the likelihood in the stacking procedure is left for future work.

This algorithm provides a path forward for broad application of shift-and-stack to large imaging surveys such as the Legacy Survey of Space and Time (LSST; \cite{Ivezić_2019}). It is very computationally challenging to perform complete image-based shift-and-stack searches on such large surveys, especially for populations of ``fast''-moving objects such as main belt asteroids and near-earth objects. As demonstrated in this work, it is possible to probe $\sim 0.5$ mag deeper into the population of main belt asteroids at relatively low computational expense by detecting at $\text{SNR}=3$ and applying this algorithm. This algorithm will be immediately applicable in finding these objects in near-ecliptic deep drilling fields of the LSST as well as in the Rubin Observatory's commissioning survey of these fields.

\begin{acknowledgments}

The authors are grateful to the anonymous reviewers for their helpful comments and suggestions which have helped improve the quality of this manuscript.

This work is supported by the National Aeronautics and Space Administration under grant No. NNX17AF21G was issued through the SSO Planetary Astronomy Program.

S.S. and M.J. acknowledge the support from the University of Washington College of Arts and Sciences, Department of Astronomy, and the DiRAC Institute. The DiRAC Institute is supported through generous gifts from the Charles and Lisa Simonyi Fund for Arts and Sciences and the Washington Research Foundation. M.J. wishes to acknowledge the support of the Washington Research Foundation Data Science Term Chair fund, and the University of Washington Provost’s Initiative in Data-Intensive Discovery.

This work made use of the following software packages: \texttt{astropy} \citep{astropy:2013, astropy:2018, astropy:2022}, \texttt{Jupyter} \citep{2007CSE.....9c..21P, kluyver2016jupyter}, \texttt{matplotlib} \citep{Hunter:2007}, \texttt{numpy} \citep{numpy}, \texttt{pandas} \citep{mckinney-proc-scipy-2010, pandas_13819579}, \texttt{python} \citep{python}, \texttt{scipy} \citep{2020SciPy-NMeth, scipy_14880408}, and \texttt{Numba} \citep{numba:2015, Numba_14713247}. Software citation information aggregated using \texttt{\href{https://www.tomwagg.com/software-citation-station/}{The Software Citation Station}} \citep{software-citation-station-paper, software-citation-station-zenodo}. 

This work was facilitated through the use of advanced computational, storage, and networking infrastructure provided by the Hyak supercomputer system at the University of Washington

This project used data obtained with the Dark Energy Camera (DECam), which was constructed by the Dark Energy Survey (DES) collaboration. Funding for the DES Projects has been provided by the US Department of Energy, the U.S. National Science Foundation, the Ministry of Science and Education of Spain, the Science and Technology Facilities Council of the United Kingdom, the Higher Education Funding Council for England, the National Center for Supercomputing Applications at the University of Illinois at Urbana-Champaign, the Kavli Institute for Cosmological Physics at the University of Chicago, Center for Cosmology and Astro-Particle Physics at the Ohio State University, the Mitchell Institute for Fundamental Physics and Astronomy at Texas A\&M University, Financiadora de Estudos e Projetos, Fundação Carlos Chagas Filho de Amparo à Pesquisa do Estado do Rio de Janeiro, Conselho Nacional de Desenvolvimento Científico e Tecnológico and the Ministério da Ciência, Tecnologia e Inovação, the Deutsche Forschungsgemeinschaft and the Collaborating Institutions in the Dark Energy Survey.

The Collaborating Institutions are Argonne National Laboratory, the University of California at Santa Cruz, the University of Cambridge, Centro de Investigaciones Enérgeticas, Medioambientales y Tecnológicas–Madrid, the University of Chicago, University College London, the DES-Brazil Consortium, the University of Edinburgh, the Eidgenössische Technische Hochschule (ETH) Zürich, Fermi National Accelerator Laboratory, the University of Illinois at Urbana-Champaign, the Institut de Ciències de l’Espai (IEEC/CSIC), the Institut de Física d’Altes Energies, Lawrence Berkeley National Laboratory, the Ludwig-Maximilians Universität München and the associated Excellence Cluster Universe, the University of Michigan, NSF NOIRLab, the University of Nottingham, the Ohio State University, the OzDES Membership Consortium, the University of Pennsylvania, the University of Portsmouth, SLAC National Accelerator Laboratory, Stanford University, the University of Sussex, and Texas A\&M University.

Based on observations at NSF Cerro Tololo Inter-American Observatory, NSF NOIRLab (NOIRLab Prop. ID 2019A-0337; PI: D. Trilling), which is managed by the Association of Universities for Research in Astronomy (AURA) under a cooperative agreement with the U.S. National Science Foundation.

\end{acknowledgments}

\bibliography{main}{}
\bibliographystyle{aasjournal}

\end{document}